\documentclass[fleqn,usenatbib]{mnras}

\usepackage{graphicx,color}	
\usepackage{graphicx}	
\usepackage{amsmath}	
\usepackage{amssymb}	

\usepackage{fancyvrb}
\usepackage{fancybox}

\definecolor{mygray}{gray}{0.7}

\title[Clustering and HOD of AGNs in $\nu^2$GC]{Semi-analytic modelling of AGNs: auto-correlation function and halo occupation}

\author[T. Oogi et al.]{
Taira Oogi,$^{1,7}$\thanks{E-mail: taira.oogi@ipmu.jp (TO)}
Hikari Shirakata,$^{2,3}$
Masahiro Nagashima,$^{4}$
Takahiro Nishimichi,$^{1,5}$
\newauthor{ Toshihiro Kawaguchi,$^{6}$ Takashi Okamoto,$^{2}$ Tomoaki Ishiyama$^{7}$
and Motohiro Enoki$^{8}$}
\\
$^{1}$Kavli Institute for the Physics and Mathematics of the Universe (WPI), The University of Tokyo Institutes for Advanced Study, \\ \ The University of Tokyo, 5-1-5 Kashiwanoha, Kashiwa, Chiba 277-8583, Japan\\
$^{2}$Department of Cosmosciences, Graduates School of Science, Hokakido University, N10 W8, Kitaku, Sapporo 060-0810, Japan\\
$^{3}$The Technical Research Center, Tadano Ltd., 2217-13, Hayashi-machi, Takamatsu, Kagawa, 761-0301, Japan\\
$^{4}$Faculty of Education, Bunkyo University, 3337 Minami-Ogishima, Koshigaya-shi, Saitama 343-8511, Japan\\
$^{5}$Center for Gravitational Physics, Yukawa Institute for Theoretical Physics, Kyoto University, Kyoto 606-8502, Japan\\
$^{6}$Department of Economics, Management and Information Science, Onomichi City University, 1600-2, Hisayamada, Onomichi, \\ \ Hiroshima 722-8506, Japan\\
$^{7}$Institute of Management and Information Technologies, Chiba University, 1-33, Yayoi-cho, Inage-ku, Chiba 263-8522, Japan\\
$^{8}$Center for General Education, Tokyo Keizai University, 1-7-34, Minami-cho, Kokubunji-shi, Tokyo 185-8502, Japan\\
}

\date{Accepted XXX. Received YYY; in original form ZZZ}

\pubyear{2019}

\begin{document}
\label{firstpage}
\pagerange{\pageref{firstpage}--\pageref{lastpage}}

\thisfancyput(14.8cm,0.5cm){\large{YITP-19-125}}

\maketitle

\begin{abstract}
The spatial clustering of active galactic nuclei (AGNs) is considered to be one of the important diagnostics for the understanding of the underlying processes behind their activities complementary to measurements of the luminosity function (LF). We analyse the AGN clustering from a recent semi-analytic model performed on a large cosmological $N$-body simulation covering a cubic gigaparsec comoving volume. 
We have introduced a new time-scale of gas accretion on to the supermassive black holes to account for the loss of the angular momentum on small scales, which is required to match the faint end of the observed X-ray LF.
The large simulation box allows us accurate determination of the auto-correlation function of the AGNs. The model prediction indicates that this time-scale plays a significant role in allowing massive haloes to host relatively faint population of AGNs,
leading to a higher bias factor for those AGNs. The model predictions are in agreement with observations of X-ray selected AGNs in the luminosity range $10^{41.5}~\mathrm{erg} \ \mathrm{s}^{-1} \leq L_{2\mathchar`-10\mathrm{keV}} \leq 10^{44.5}~\mathrm{erg} \ \mathrm{s}^{-1}$, with the typical host halo mass of $10^{12.5-13.5} h^{-1}\,{\rm M}_{\odot}$ at $z \la 1$.
This result shows that the observational clustering measurements impose an independent constraint on the accretion time-scale complementary to the LF measurements.
Moreover, we find that not only the effective halo mass corresponding to the overall bias factor, but the extended shape of the predicted AGN correlation function shows remarkable agreement with those from observations. Further observational efforts towards the low luminosity end at $z \sim 1$ would give us stronger constraints on the triggering mechanisms of AGN activities through their clustering.
\end{abstract}

\begin{keywords}
galaxies: haloes -- galaxies: nuclei -- quasars: general -- dark matter -- large-scale structure of Universe -- cosmology: theory
\end{keywords}



\section{Introduction}

Observations have shown that there are a number of scaling relations between the mass of supermassive black holes (SMBHs) and the properties of their host galaxies.
The SMBH mass correlates with the bulge stellar mass of the host galaxy (e.g. \citealt{Magorrian1998}; \citealt{Haring2004}; \citealt{McConnell2013}) and with the stellar velocity dispersion (e.g. \citealt{Ferrarese2000}; \citealt{Gebhardt2000}; \citealt{McConnell2013}), although the nature of these relationships for low-mass galaxies is still poorly understood (see e.g. \citet{Graham2015}, \citet{Chilingarian2018}, and \citet*{Schutte2019} for recent studies).
These correlations suggest that the growth of SMBHs and the evolution of the host galaxies are physically related.

While there are many studies on physical processes, which can potentially connect the cosmic growth of the SMBHs and the host galaxies, it is still unclear how exactly their correlation has been established over cosmic time.
A scenario invoking galaxy mergers is popular for the growth of SMBHs because a merger can stimulate large gas inflow to the central region of galaxies and feed the SMBHs owing to the gravitational torques induced by the interaction.
This is indeed observed in hydrodynamical simulations, powering active galactic nucleus (AGN) as well as triggering starbursts in the nuclear region (e.g. \citealt*{DiMatteo2005Natur.433..604D}; \citealt{Hopkins2006ApJS..163....1H}).
Observational studies are consistent with the galaxy merger scenario at least for a subset of quasar samples (i.e., high-luminosity AGNs, \citealt{Bessiere2012}; \citealt{Urrutia2012}; \citealt{Treister2012}; \citealt{Glikman2015}; \citealt{Fan2016}).
On the other hand, it is also expected that gas inflow due to the gravitational instability of self-gravitating (gas) discs causes the gas accretion on to SMBHs even in the absence of mergers.
This process is called disc instabilities (e.g. \citealt{Bower2006}; \citealt{Lagos2008}; \citealt{Fanidakis2011}; \citealt{Hirschmann2012}; \citealt{Menci2014}; \citealt{Gatti2016}; \citealt{Lacey2016}; \citealt{Griffin2019}).
This scenario is supported by observational studies investigating the morphology of galaxies hosting AGNs with moderate X-ray luminosity.
In this paper, $L_{\rm X}$ denotes
the luminosity of 2-10 keV band, unless otherwise stated.
Galaxies hosting moderately luminous AGNs ($L_{\rm X}\approx 10^{42} \mathchar`- 10^{44}~\mathrm{erg~s^{-1}}$) do not show evidence of morphological distortion rooted in major galaxy mergers at $z<1.3$ (e.g. \citealt{Grogin2005}; \citealt{Gabor2009}; \citealt{Georgakakis2009}; \citealt{Cisternas2011}) and at $z\sim2$ (\citealt{Schawinski2011}; \citealt{Kocevski2012}).
More recent observational studies also suggest that the merger fraction of AGN host galaxies is similar to that of inactive galaxies for luminous AGNs (e.g. \citealt{Mechtley2016}; \citealt{Marian2019}).
Note that different observations use different X-ray bands including 2-8, 2-10, 0.5-10, 0.2-10, and 0.5-8 keV bands.
Analyses using close galaxy pairs find that although $\sim 20$ percent of AGN activity is triggered by close encounters or mergers between galaxies, the triggering mechanism for the remaining $\sim80$ percent is open to question (e.g. \citealt{Silverman2011}; \citealt{Lackner2014}).
The merger-AGN connection is, however, still of interest and remains a matter of debate. This is because a number of studies suggest that galaxies undergoing mergers or interacting with pairs have a higher fraction of AGN compared to undisturbed galaxies (e.g. \citealt{Satyapal2014}; \citealt{Goulding2018}; \citealt{Ellison2019}; \citealt{Secrest2020}).

The similarity between the cosmic star formation history and the SMBH accretion history (e.g. \citealt{Silverman2008}; \citealt{Aird2010}; \citealt{Madau2014ARA&A}) as well as the correlations between the SMBH mass and the bulge mass described above (\citealt{Magorrian1998}; \citealt{Haring2004}; \citealt{McConnell2013}) suggest 
a possible connection between the growth of SMBHs and the formation and evolution of galaxies in the cosmological structure formation.
The evolution of AGN populations over cosmic time is a key topic of observational and theoretical studies, and clarifying it has important meaning in theoretical models of AGNs.
Semi-analytic models of galaxy formation including the processes of the growth of SMBHs and AGN activity, which are driven by mergers and/or disc instabilities, have been developed to study the cosmological evolution of SMBHs and galaxies (e.g. \citealt{Kauffmann2000}; \citealt*{Enoki2003}; \citealt{Cattaneo2005}; \citealt{Fontanot2006}; \citealt*{Monaco2007}; \citealt{Marulli2008}; \citealt{Somerville2008}; \citealt{Fanidakis2011, Fanidakis2012}; \citealt{Hirschmann2012}; \citealt*{Menci2013}; \citealt{Neistein2014}; \citealt{Menci2014}; \citealt{Enoki2014}; \citealt{Shirakata2019}; \citealt{Griffin2019}).
These studies mainly focus on the AGN luminosity function over a wide redshift range.
Although recent semi-analytic models have predicted the AGN luminosity function which is in agreement with observations (e.g. \citealt{Fanidakis2012}; \citealt{Hirschmann2012}; \citealt{Shirakata2019}; \citealt{Griffin2019}), the main triggering mechanism which shapes the AGN luminosity function is different among the models.
Consequently, the main contributor to the SMBH growth remains unclear.

Spatial clustering of AGNs also gives an important constraint on the SMBH growth and AGN triggering mechanisms.
Large-scale surveys such as the Sloan Digital Sky Survey (SDSS; \citealt{York2000}) and the 2dF QSO Redshift Survey (2QZ; \citealt{Boyle2000}) have revealed that for optically luminous quasars the host halo mass inferred from clustering analysis is a few times $10^{12} h^{-1}\,{\rm M}_{\odot}$ (e.g. \citealt*{Porciani2004}; \citealt{Croom2005}; \citealt{Myers2007}; \citealt{Shen2007}; \citealt{Padmanabhan2009}; \citealt{Ross2009}; \citealt{Krolewski2015}).
On the other hand, X-ray AGN surveys have found that, for moderately luminous X-ray-selected AGNs ($L_{2-10\mathrm{keV}}\simeq 10^{42} \mathchar`- 10^{44}~\mathrm{erg~s^{-1}}$), the host halo mass must be higher, typically $10^{12.5-13.5} h^{-1}\,{\rm M}_{\odot}$, to be compatible with their clustering measures (e.g. \citealt{Gilli2009}; \citealt{Coil2009}; \citealt*{Krumpe2010}; \citealt{Cappelluti2010}; \citealt{Miyaji2011}; \citealt{Allevato2011}; \citealt{Starikova2011}; \citealt{Mountrichas2012}; \citealt{Koutoulidis2013}; \citealt{Krumpe2018}; \citealt{Plionis2018}).
It is expected that next generation surveys such as eROSITA (\citealt{Merloni2012}) can reveal the clustering of X-ray AGNs in a wide luminosity range more accurately.

Previous theoretical studies have indicated that at least at low redshift ($z\la2$) the dark matter (DM) halo mass of luminous quasars can be explained by the models in which quasar activity is triggered by galaxy major mergers (e.g. \citealt{Hopkins2007ApJ...662..110H}; \citealt{Bonoli2009}; \citealt{Oogi2016}) or disc instabilities (\citealt{Fanidakis2013b}).
In order to explain the X-ray AGN host haloes, some physical mechanisms are proposed.
\citet{Fanidakis2013a} compare the theoretical predictions of the host halo mass of X-ray AGNs based on a semi-analytic model to the observational estimates.
In their model, SMBHs grow through the accretion of diffuse hot gas from a quasi-hydrostatic halo as well as the accretion of cold gas during starbursts.
The former `hot-halo' accretion mode is regarded as being coupled to the AGN feedback.
They further take into account the AGN activity associated with this hot-halo mode accretion and add it to the AGN luminosity.
Because the AGN feedback in the quasi-hydrostatic regime is activated in massive haloes ($\ga 10^{12.5} {\rm M}_{\odot}$), they are more likely to be found in those haloes.
As a consequence, they have shown that the higher host halo mass of the X-ray AGNs can be explained by the hot-halo mode AGNs.
They have suggested that the difference between the host halo masses of quasars and X-ray AGNs is due to the difference in black hole fuelling/AGN triggering modes between the two populations (\citealt{Fanidakis2013a}).
However, it is still unclear whether the hot-halo mode accretion is the only solution as the major triggering mechanism of X-ray AGNs or not.

Other possibilities that may trigger the X-ray AGNs have been also considered.
\citet{Allevato2011} suggest that for moderate-luminosity X-ray selected AGNs (with the bolometric luminosity $L_{\mathrm{bol}} \sim 2 \times 10^{45}~\mathrm{erg} \ \mathrm{s}^{-1}$) secular processes such as tidal disruptions or disc instabilities might play a larger role than major mergers, based on their observation.
\citet{Altamirano-Devora2016} estimate the halo occupation distributions (HODs) of the moderate-luminosity X-ray AGNs ($L_{0.5\mathchar`-2 \mathrm{keV}} \geq 2 \times 10^{42.4}~h^{-2}\,\mathrm{erg} \ \mathrm{s}^{-1}$) using a simple model with cosmological $N$-body simulations and compare them with those inferred from observations.
They suggest that minor mergers play a role in establishing the HODs and can be an important factor in activating X-ray AGNs.
\citet{Gatti2016} use their semi-analytic model to clarify the difference of predictions between two AGN triggering mechanisms: galaxy interactions and disc instabilities.
They have shown that both scenarios with galaxy interactions alone or with only disc instabilities do not sufficiently reproduce the HODs by using direct measurements by directly counting the number of AGNs within X-ray galaxy groups (\citealt{Allevato2012}) or using an abundance-matching approach (\citealt{Leauthaud2015}) for X-ray AGNs, in particular, the HODs of satellite AGNs.
They have concluded that different feeding modes besides galaxy interactions including mergers and disc instabilities are needed to account for the triggers of satellite AGNs.

In this paper, we investigate the clustering of X-ray AGNs by using our new semi-analytic model $\nu^2$GC (\citealt{Makiya2016}; \citealt{Shirakata2019}).
In particular, we explore the validity of $\nu^2$GC including the galaxy major and minor mergers and disc instabilities  as triggering mechanisms for X-ray AGNs.
In $\nu^2$GC, we assume that gas accretion on to SMBHs occurs when two galaxies merge or a galactic disc is dynamically unstable.
In these phases, we also assume that nuclear starburst occurs.
Furthermore we assume that the time-scale of the BH growth, i.e. the gas accretion, is
controlled by the time-scale of the loss of angular momentum within 100~pc of the nucleus.
The model including this additional time-scale has been shown to match well the faint end of the observed luminosity functions of X-ray AGNs and reproduce those in a wide redshift range $0 \la z \la 6$ (\citealt{Shirakata2019}).
Moreover, the model predicts the Eddington ratio distribution, which is broadly consistent with observational estimates at low redshift (\citealt{Shirakata2019b}).
In this paper, we focus more on the clustering properties of the AGNs with a particular focus on the role of the accretion time-scale to determine the mass of the host haloes of X-ray AGNs.
More specifically, we examine the two-point correlation function (2PCF) and the derived bias factor of the simulated AGNs in comparison to observations.
For this purpose, we use a large cosmological $N$-body simulation covering a cubic gigaparsec comoving volume with sufficient mass resolution.
This enables us to examine 2PCFs with unprecedentedly high accuracy.
We find that the effective halo mass of X-ray AGNs increases with cosmic time and reaches $10^{12.5-13.5} h^{-1}\,{\rm M}_{\odot}$ at $0\la z \la 1$ in our model and show that the result is consistent with the observationally estimated host halo mass of X-ray AGNs.

The reminder of this paper is organized as follows.
In Section 2, we outline our semi-analytic model of galaxy and AGN formation to predict the AGN statistics.
Our primary results including the host halo masses, 2PCFs, and halo occupation distributions of AGN are presented in Section 3.
In Section 4, we discuss the implications of our results for the SMBH growth and AGN formation processes, and summarize our findings.

\section{Methods}
\label{sec:methods}

\subsection{Semi-analytic galaxy formation model}

To investigate the clustering of AGNs, we use our latest semi-analytic model of galaxy and AGN formation,  `{\it New Numerical Galaxy Catalogue}', $\nu^2$GC (\citealt{Makiya2016}; \citealt{Shirakata2019}), which is an extension of the Numerical Galaxy Catalog ($\nu$GC; \citealt{Nagashima2005}).
DM halo merger trees are constructed by using large cosmological $N$-body simulations (\citealt{Ishiyama2015}).
In this paper, we use the simulation run called $\nu^2$GC-L, which has the largest box size, $1120.0~h^{-1}\,\mathrm{Mpc}$, among the different runs in the $\nu$GC project, while keeping sufficient mass resolution for the study of AGNs.
The mass of the DM particles is $2.2 \times 10^{8}~h^{-1}\,{\rm M}_{\odot}$, and the total number of particles is $8192^3$.
We consider DM haloes with mass equal to or more massive than $8.79\times10^{9}~h^{-1}\,{\rm M}_{\odot}$, which consists of 40 DM particles.
The semi-analytic model on $\nu^2$GC is originally developed by \citet{Makiya2016} and extended by \citet{Shirakata2019} with several major modifications including the growth of SMBHs and bulges and the AGN activity.
$\nu^2$GC takes into account all the known major processes involved in galaxy formation: (i) the collapse and merging of DM haloes, (ii) radiative gas cooling and disc formation in DM haloes, (iii) star formation, supernova feedback and chemical enrichment, (iv) galaxy mergers and disc instabilities, which trigger starburst and cause the growth of bulges and SMBHs, and (v) feedback from AGNs (the so-called radio-mode AGN feedback).
Further details of our galaxy formation model are given in \citet{Makiya2016} and \citet{Shirakata2019}.

\subsection{Bulge and SMBH growth}
In $\nu^2$GC, there are two main channels to feed the SMBHs: galaxy mergers and disc instabilities (\citealt{Shirakata2019}).
These fueling mechanisms are assumed to induce a starburst in the nuclear region of galaxies.
The newly formed stars constitute the bulge component of the galaxies.
The growth of the SMBHs is considered to follow that of the bulges.
First, we describe the merger-driven growth mode.
This mode is motivated by a number of galaxy merger simulations (e.g. \citealt{DiMatteo2005Natur.433..604D}; \citealt{Springel2005ApJ...620L..79S}; \citealt{Springel2005MNRAS.361..776S}; \citealt{Hopkins2006ApJS..163....1H}).
Right after the mergers of DM haloes, the newly formed halo contains two or more galaxies.
The central galaxy in the most massive progenitor halo is defined as that of the new halo.
The others are defined as satellite galaxies.
We assume that two galaxies sharing a common DM halo merge into a single galaxy in a merger time-scale, which is determined by the dynamical friction time-scale (\citealt{Jiang2008}; \citealt{Jiang2010}) and/or the rate of random collision (\citealt{Makino1997}).
We consider two different types of mergers: (i) a merger between a galaxy at the halo centre and a satellite galaxy in the same halo in the dynamical friction time-scale and (ii) a merger between two satellite galaxies according to the random collision rate.
We follow the model of merger-driven bulge growth proposed by \citet{Hopkins2009ApJ...691.1168H} based on hydrodynamical simulations of galaxy mergers.
Through the galaxy mergers, all the components of the secondary (less massive) galaxy merge into the bulge of the primary (more massive) galaxy.
A part of the cold gas and stars in the disc of the primary galaxy also migrates to the bulge due to the interaction.
The cold gas is converted into stars in the bulge with a short time-scale, which we call a starburst.
During starbursts, a small fraction of the cold gas is considered to be accreted on to the SMBHs.
We note that the starbursts and the accompanying gas accretion on to SMBHs occur both in major and minor mergers in our model.
We define mergers with mass ratios of the secondary to the primary larger (smaller) than 0.4 as major (minor) mergers.
This definition is the same as in \citet{Shirakata2019}.
Further details of the prescription of the mergers are given in \citet{Makiya2016} and \citet{Shirakata2019}.

We also consider the SMBH growth driven by disc instabilities.
When a galactic disc is dynamically unstable, a part of the cold gas and stars in the disc migrates to the bulge through the bar formation.
This process is supported by numerical simulations of isolated galaxies (e.g. \citealt*{Efstathiou1982}) and an analytic model (\citealt{Hopkins2011MNRAS.415.1027H}).
We follow the stability criterion given by \citet{Efstathiou1982}.
The starbursts triggered by disc instabilities are treated in the same way as those by mergers.
Here again, a fraction of the gas is considered to be accreted on to the SMBHs.
Further details of our bulge and SMBH growth model are given in \citet{Shirakata2019}.

We also assume that a fraction of hot halo gas is accreted on to SMBHs.
In this mode, the accreted gas causes a radio jet that puts energy into the hot halo gas and prevents the hot gas from cooling and subsequent star formation.
This is called the `radio-mode' AGN feedback (\citealt{Croton2006}; \citealt{Bower2006}).
Although SMBHs grow through this gas accretion mode, this is not a significant contribution to the entire mass growth of SMBHs; the mass growth is dominated by the gas accretion during major/minor mergers and disc instabilities.
We note that our model does not consider the AGN luminosity associated with this hot halo gas accretion, in contrast to \citet{Fanidakis2013a}.
Further details of the radio-mode feedback we implement are given in \citet{Makiya2016} and \citet{Shirakata2019}.

\subsection{Gas accretion time-scales on to black holes}

When the triggering mechanisms are in action, the gas moves toward the central region of the galaxy and is accreted on to the SMBH.
The accreted gas mass $\Delta M_{\mathrm{BH}}$ is modelled as follows:
\begin{equation}
\Delta M_{\mathrm{BH}} = f_{\mathrm{BH}} \Delta M_{\mathrm{star,burst}},
\end{equation}
where $\Delta M_{\mathrm{star,burst}}$ is the total stellar mass formed during a starburst.
We set $f_{\mathrm{BH}}= 0.02$ to match the observed correlation between masses of the host bulges and the SMBHs at $z=0$.
This model also reproduces the SMBH mass function estimated from observations.
We assume a time-dependent mass accretion associated with each event:
\begin{equation}
\dot{M}_{\mathrm{BH}} (t) = \frac{\Delta M_{\mathrm{BH}}}{t_{\mathrm{acc}}}\exp \left(- \frac{t-t_{\mathrm{start}}}{t_{\mathrm{acc}}}\right),
\end{equation}
where $t_{\mathrm{acc}}$ is the accretion time-scale and $t_{\mathrm{start}}$ is the start time of the accretion.

The accretion time-scale, $t_{\mathrm{acc}}$, is a key parameter that governs the abundance of low-luminosity AGNs.
In \citet{Shirakata2019}, we have shown that a long accretion time-scale is necessary to reproduce the faint end of the AGN luminosity function.
We assume that the this time-scale is controlled by that of the loss of angular momentum within 100~pc of a galaxy in addition to the dynamical time-scale of the bulge:
\begin{equation}
t_{\mathrm{acc}} = \alpha_{\mathrm{bulge}} t_{\mathrm{dyn, bulge}} + t_{\mathrm{loss}},
\end{equation}
where $t_{\mathrm{dyn, bulge}}$ is the dynamical time-scale of the bulge of the host galaxy, $\alpha_{\mathrm{bulge}}$ is a free parameter, and $t_{\mathrm{loss}}$ is the time-scale for the angular momentum loss of the accreted gas within 100~pc \citep{Shirakata2019}.
We set $\alpha_{\mathrm{bulge}} = 0.58$ so that the bright end of the luminosity function of AGNs matches the observations.
In this prescription, if $t_{\mathrm{loss}}$ is sufficiently longer than $\alpha_{\mathrm{bulge}} t_{\mathrm{dyn, bulge}}$, the gas accretion continues even after passing the bulge dynamical time-scale, which is assumed to be the time-scale of the gas supply from the galactic scale.
Here, we describe the motivation to introduce the new time-scale $t_{\mathrm{loss}}$ and the situation we suppose.
We consider that the gas accretion on to SMBHs is regulated by the physics that governs the dynamics of gas around the SMBHs.
The dynamics of a circumnuclear disc and/or accretion disc is considered to be related to the SMBH mass and the accreted gas mass.
Thus, we assume that $t_{\mathrm{loss}}$ depends both on the SMBH mass, $M_{\mathrm{BH}}$, and $\Delta M_{\mathrm{BH}}$:
\begin{equation}
t_{\mathrm{loss}} = t_{\mathrm{loss,0}} \left( \frac{M_{\mathrm{BH}}}{{\rm M}_{\odot}} \right)^{\gamma_{\mathrm{BH}}} \left( \frac{\Delta M_{\mathrm{BH}}}{{\rm M}_{\odot}} \right)^{\gamma_{\mathrm{gas}}},
\end{equation}
where $t_{\mathrm{loss,0}}$, $\gamma_{\mathrm{BH}}$, and $\gamma_{\mathrm{gas}}$ are free parameters.
We set $t_{\mathrm{loss,0}} = 1$~Gyr, $\gamma_{\mathrm{BH}} = 3.5$, and $\gamma_{\mathrm{gas}} = -4.0$ so that the AGN luminosity function matches observations.
The positive dependence on $M_{\mathrm{BH}}$ and negative dependence on $\Delta M_{\mathrm{BH}}$ is in accord with the situation that we consider: a massive SMBH stabilizes the gas disc due to its gravitational potential, and a massive accreted gas mass makes the gas disc unstable.
For most of the AGNs with low-luminosity and/or at low redshifts, the dominant term in $t_{\mathrm{acc}}$ is the second term $t_{\mathrm{loss}}$ as shown in \citet{Shirakata2019} (see their fig.~7).
Due to this term, the number density of AGNs with low luminosities increases at $z\la 1.5$.
As a result, this model can reproduce the observed luminosity function of AGNs over a wide redshift range, $0<z<6$ (see fig.~5 in \citealt{Shirakata2019}).
We adopt the model with $t_{\mathrm{loss}}$ as our fiducial model.
To clarify the effect of $t_{\mathrm{loss}}$, we also consider the model without the term $t_{\mathrm{loss}}$ in $t_\mathrm{acc}$, which we call `galmodel'.
In this model, we assume that the accretion time-scale only depends on the dynamical time-scale of the bulge, i.e. $t_{\mathrm{acc}} = \alpha_{\mathrm{bulge}} t_{\mathrm{dyn, bulge}}$.

\subsection{AGN activity}

The gas inflow and accretion on to a SMBH leads to AGN activities.
Here, we consider the AGN bolometric luminosity, $L_{\mathrm{bol}}$, as the most representative indicator of these activities.
When the accretion occurs at a sub-Eddington rate, i.e. $\dot{M}_{\mathrm{BH}} \la \dot{M}_{\mathrm{Edd}} / \eta$, the accretion disc is described as the standard disc, where $\dot{M}_{\mathrm{Edd}}$ is the Eddington rate and $\eta$ is the radiative efficiency of a standard disc, $\eta \sim 0.1$, respectively.
The Eddington rate is defined as $\dot{M}_{\mathrm{Edd}} \equiv L_{\mathrm{Edd}}/c^2$, where $L_{\mathrm{Edd}}$ is the Eddington luminosity $L_{\mathrm{Edd}} \equiv 4\pi G M_{\mathrm{BH}} m_{p} c / \sigma_{\scalebox{0.5}{$T$}}$, $c$ is the speed of light, $G$ is the gravitational constant, $m_{p}$ is the proton mass, and $\sigma_{\scalebox{0.5}{$T$}}$ is the Thomson scattering cross-section.
When a super-Eddington accretion occurs, the luminosity is expected to be limited to several times the Eddington luminosity due to the `photon trapping' (e.g. \citealt{Ohsuga2005}).
In other words, the radiative efficiency is lower than that of the standard accretion disc given the accretion rate.
This type of disc state is described by a solution called slim disc (e.g. \citealt{Abramowicz1988}; \citealt{Watarai2000}; \citealt{Mineshige2000}).
\citet{Oogi2017} have found that this limitation of the bolometric luminosity has a large impact on the clustering signal of the resultant mock AGN population.
To include this effect, we adopt the following relation as the bolometric luminosity normalised by the Eddington luminosity, $\lambda_{\mathrm{Edd}} \equiv L_{\mathrm{bol}} / L_{\mathrm{Edd}}$:
\begin{equation}
\lambda_{\mathrm{Edd}} = \left[ \frac{1}{1+3.5 \{ 1+\tanh(\log (\dot{m} / \dot{m}_{\mathrm{crit}})) \} }  + \frac{\dot{m}_{\mathrm{crit}}}{\dot{m}}  \right]^{-1},
\end{equation}
where $\dot{m}$ is the accretion rate normalised by the Eddington rate $\dot{M}_{\mathrm{Edd}}$, $\dot{m} \equiv \dot{M}_{\mathrm{BH}} / \dot{M}_{\mathrm{Edd}}$, and
the parameter $\dot{m}_{\mathrm{crit}}$ should be in the range, $2.5 \la \dot{m}_{\mathrm{crit}} \la 16.0$.
This form is based on \citet{Kawaguchi2003}.
We set $\dot{m}_{\mathrm{crit}}=10.0$.
In this case, $\lambda_{\mathrm{Edd}}$ monotonically increases with approximately constant slope, $d\lambda_{\mathrm{Edd}} / d\dot{m} = 0.1$, until $\dot{m} \sim 30$.
The transition of the disc state to the slim disc slows down the increase of $\lambda_{\mathrm{Edd}}$, and $\lambda_{\mathrm{Edd}}$ rises to the maximum value, $\lambda_{\mathrm{Edd}}=8$, at $\dot{m} \rightarrow \infty$ (see also fig.~1 in \citealt{Shirakata2019b}).
This dependence on $\dot{m}$ is similar to the models in \citet{Watarai2000} and \citet{Mineshige2000}.
Further, we use the bolometric correction from \citet{Marconi2004} to obtain the hard X-ray (2-10~keV) luminosity of AGNs.
We have investigated the effects of the different bolometric correction on the result, in particular, the two-point correlation function.
We use the \citet{Netzer2019} bolometric correction for comparison.
In this case, $L_{2\mathchar`-10\mathrm{keV}}$ is about a factor of two larger than that assuming the \citet{Marconi2004} bolometric correction for a given bolometric luminosity.
We find that, despite such a large impact on the luminosity, the 2PCF is not significantly altered by the choice of different bolometric corrections, well below the statistical uncertainties of the observational data.

\subsection{Clustering analysis}

We here describe how we measure the two-point correlation function (2PCF) of AGNs.
The large-scale cosmological $N$-body simulations we use enable us to calculate
the 2PCF with high accuracy.
We first estimate the two-dimensional 2PCF of AGNs, $\xi (r_p, \pi)$ using an estimator:
\begin{equation}
\xi (r_p, \pi) = \frac {\mathrm{AA} (r_p, \pi)} {\mathrm{RR} (r_p, \pi)} - 1,
\end{equation}
where $r_p$ and $\pi$ are the separations perpendicular and parallel to the line of sight,
$\mathrm{AA} (r_p, \pi)$ and $\mathrm{RR} (r_p, \pi)$ are AGN-AGN and random-random pairs at a given separation, $r_p$ and $\pi$, respectively.
In terms of $\pi$, we do not take into account the contamination from peculiar velocities in the redshift.
The pair counts are normalised by the total number of pairs of AGNs and random particles, respectively.
We then compute the projected correlation function, $w_p (r_p)$, by
\begin{equation}
\label{eq:wp}
w_p (r_p) = 2 \int_0^{\pi_{\mathrm{max}}} \xi (r_p, \pi) \mathrm{d}\pi,
\end{equation}
where we use $\pi_{\mathrm{max}} = 40~h^{-1}$~Mpc as the default value for our predictions. We also consider other values for $\pi_\mathrm{max}$ for a fair comparison with observational data.
For $\pi_{\mathrm{max}} \geq 40~h^{-1}$~Mpc, we have confirmed that neglecting the peculiar velocities does not affect the amplitude of $w_p (r_p)$ significantly (by only 5 per cent, or 0.1 dex in halo mass).
When we choose $\pi_{\mathrm{max}} = 10~h^{-1}$~Mpc (see Section 3.3 and Fig.~\ref{fig:2pcf_obs}), the amplitude of the 2PCF with the peculiar velocity increases in the large-scale separation and matches the observation better.
We postpone further treatment of the peculiar velocity in future studies.

We estimate the statistical errors of the 2PCF using the jackknife resampling method (e.g. \citealt{Krumpe2010}).
We divide the simulation volume into sixty-four sub-volumes and calculate $w_p (r_p)$ sixty-four times, where each jackknife sample excludes one sub-volume.
These 2PCFs are used to derive the covariance matrix $M_{ij}$ by
\begin{equation}
\begin{split}
M_{ij} = \frac{N_{\mathrm{run}} - 1} {N_\mathrm{run}} \left[ 
\sum_{k=1}^{N_\mathrm{run}} (w_{p,k}(r_{p,i}) - \langle w_p(r_{p,i}) \rangle ) \right.\\
\times (w_{p,k}(r_{p,j}) - \langle w_p(r_{p,j}) \rangle ) 
\Bigg],
\end{split}
\end{equation}
where $w_{p,k} (r_{p,i})$ is the 2PCF value for the $k$-th jackknife sample and $\langle w_p(r_{p,i}) \rangle$ is the average over all of the jackknife samples.
$N_\mathrm{run}$ is the number of the jackknife samples and we adopt $N_\mathrm{run}=64$ in this analysis.
The covariance matrix $M_{ij}$ reflects the degree to which the $i$-th bin
of $r_p$ is correlated with the $j$-th bin.
The 1$\sigma$ error of each $r_p$ bin is given by the square root of the diagonal component of $M_{ij}$, $\sigma_i = \sqrt{M_{ii}}$, which is plotted as the error bars of the measured 2PCF in the figures in what follows.

In this paper, we compute the bias factor of AGNs through the AGN projected 2PCF.
We define the AGN bias as the square root of the ratio of the projected 2PCF of the AGNs to that of DM:
\begin{equation}
\label{eq:bias}
b_{\mathrm{AGN}} (r_p) = \sqrt{ \frac{w_{p, \mathrm{AGN}} (r_p)}{w_{p, \mathrm{DM}} (r_p)} }.
\end{equation}
We average this function from 1 to 20~$h^{-1}$~Mpc at different bins of $r_p$.
We mainly report this average as the representative value of bias.
This definition of the bias factor is different from that in \citet{Fanidakis2013a},
in which it is computed by weighting the bias of DM haloes by the halo occupation
distribution of AGNs, $N_{\mathrm{AGN}} (M)$.
The bias derived with this procedure is called the effective bias (\citealt{Baugh1999}).
We investigate the difference between the two bias values in Section \ref{subsec:b_comparison}.

We also consider the AGN bias measured through a power-law fit to $w_{p, \mathrm{AGN}} (r_p)$. In actual situations, this approach is often employed.
We follow this convention and provide the resultant bias values for a fair comparison to those obtained from observations in the literature.
If the auto-correlation function is expressed as $\xi (r) = (r/r_0)^{-\gamma}$, then $w_{p, \mathrm{AGN}} (r_p)$ can be directly related to $\xi (r)$ by
\begin{equation}
\label{eq:wp_power_law}
w_p (r_p) = r_p \left( \frac{r_0}{r_p} \right)^{\gamma} \frac{\Gamma(1/2) \Gamma[(\gamma-1)/2]}{\Gamma(\gamma/2)},
\end{equation}
where $\Gamma(x)$ is the Gamma function.
We fit $w_{p, \mathrm{AGN}} (r_p)$ with Eqn.~\ref{eq:wp_power_law} with $\gamma$ and $r_0$ as free parameters,
by minimizing $\chi^2$ taking into account the covariance matrix (e.g. \citealt{Miyaji2007}; \citealt{Krumpe2010}):
\begin{equation}
\begin{split}
    \chi^2 = \sum_{i=1}^{N_\mathrm{bin}} \sum_{j=1}^{N_\mathrm{bin}} 
    (w_p(r_{p,i}) - w_p^{\mathrm{model}}(r_{p,i})) \\
    \times M_{ij}^{-1} (w_p(r_{p,j}) - w_p^{\mathrm{model}}(r_{p,j})).
\end{split}
\end{equation}
We fit the data in a range $r_p = 1\mathchar`-20~h^{-1}$~Mpc, as most observational studies adopt similar ranges.
Table \ref{tbl:fitting_results} shows the results of the power-law fit.
We further consider another definition of the AGN bias using the relation
\begin{equation}
\label{eq:bias_power_law}
b_{\mathrm{AGN}} = \frac{\sigma_{8, \mathrm{AGN}} (z)} {\sigma_8 (z)},
\end{equation}
where $\sigma_{8, \mathrm{AGN}} (z)$ and $\sigma_8 (z)$ are the rms fluctuations of AGN and DM density distribution within a sphere with a comoving radius of 8~$h^{-1}$~Mpc.
For a power-law correlation function, $\sigma_{8, \mathrm{AGN}}$ can be calculated by (\citealt{1980lssu.book.....P}; \citealt{Miyaji2007})
\begin{equation}
(\sigma_{8, \mathrm{AGN}})^2 = J_2(\gamma) \left( \frac{r_0}{8~h^{-1}~\mathrm{Mpc}} \right)^{\gamma},
\end{equation}
where $J_2(\gamma) = 72 / [(3-\gamma)(4-\gamma)(6-\gamma)2^{\gamma}]$.
We derive $\sigma_{8, \mathrm{AGN}}$ from the best-fitting parameters of the power-law fits to $w_{p, \mathrm{AGN}} (r_p)$ for each redshift.
The rms linear dark matter fluctuation $\sigma_8 (z)$ can be calculated by $\sigma_8 (z) = D(z) \sigma_8 (z=0)$, where $D(z)$ is the linear growth factor normalised to 1 at $z=0$.
The uncertainty in $\sigma_{8, \mathrm{AGN}}$ is computed from those in $\gamma$ and $r_0$ by finding the values at which $\chi^2 = \chi_{\mathrm{min}}^2 + 1.0$ (\citealt{Krumpe2010, Krumpe2012}).
This is then propagated to the uncertainty on $b_\mathrm{AGN}$.
The derived values of $b_\mathrm{AGN}$ are listed in Table \ref{tbl:fitting_results}.

Finally, we introduce the effective halo mass, $M_{\mathrm{halo,eff}}$, defined as the mass which satisfies
$b_{h} (M_{\mathrm{halo,eff}}) = b_{\mathrm{AGN}}$, where we use the functional form of $b_{h} (M_{\mathrm{halo}})$ calibrated by \citet{Tinker2010}.
The derived values of $M_{\mathrm{halo,eff}}$ are listed in Table \ref{tbl:fitting_results}.
This effective halo mass corresponds to  
the typical halo mass for the hosts of AGNs, and is exact value when all the AGNs are in haloes with the same mass.
In \citet{Oogi2016}, we have confirmed that the effective halo mass calculated from the quasar bias is similar to the median of the distribution of their host halo mass.
It is, however, not fully clear if the same is true for the latest semi-analytic model, which includes major and minor mergers of galaxies as well as disc instabilities as triggering mechanisms of AGNs.
Actually, in \citet{Leauthaud2015}, the effective halo mass has been found to deviate significantly from the median of the host halo mass distribution for a sample of obscured X-ray AGNs with $10^{41.5}~\mathrm{erg} \ \mathrm{s}^{-1} < L_{0.5\mathchar`-10\mathrm{keV}} < 10^{43.5}~\mathrm{erg} \ \mathrm{s}^{-1}$ at $z<1$.
We examine the relation between the effective halo mass and the median halo mass quantitatively using our latest model in Section 3.4.

\section{Results}
\label{sec:result}

\begin{figure*}
    \begin{center}
      \includegraphics[width=170mm]{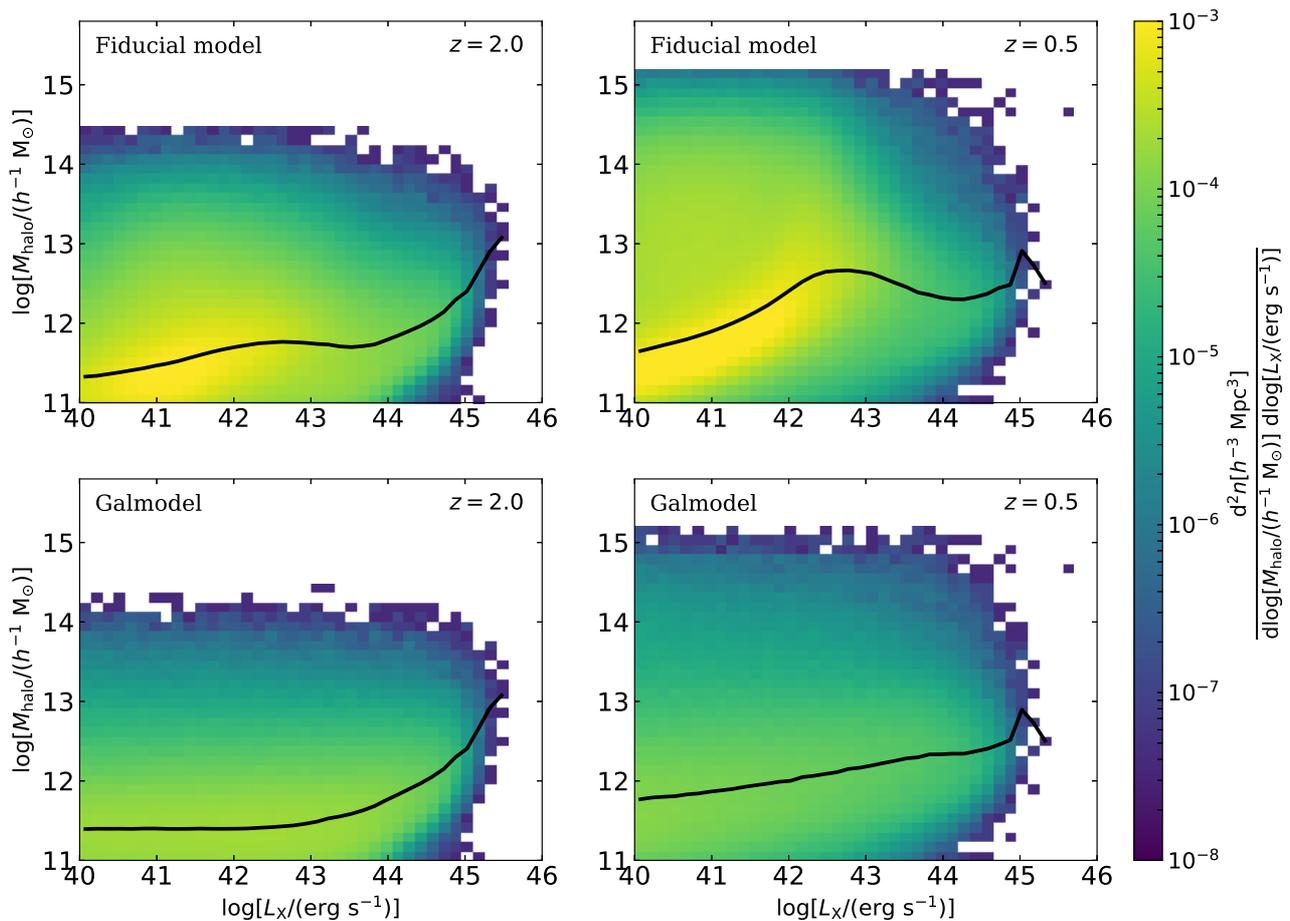}
    \end{center}
  \caption{Relations between AGN X-ray luminosity and the host DM halo mass of our model AGNs at $z=2.0$ (left) and $z=0.5$ (right).
  The color scale depicts the number density of AGNs per pixel.
  The predictions of our fiducial model are top panels.
  Those of galmodel, in which we do not take into account the gas accretion time-scale $t_{\mathrm{loss}}$, are bottom panels.
  The solid line in each panel depicts the median of the halo mass in each luminosity bin.
  }
  \label{fig:lx_mhalo_4panel}
\end{figure*}

\subsection{$L_{\rm X}-M_{\mathrm{halo}}$ relation}
\label{subsec:lx_mhalo}

Firstly, we show the relation between the X-ray luminosity of AGNs and their host DM halo mass predicted by our model in Fig.~\ref{fig:lx_mhalo_4panel}, which gives us some insights into the AGN clustering, especially its luminosity and redshift dependence. The plot shows the number density of the AGNs in the unit log-interval of luminosity and mass by as shown by the color bar.
Overall, the luminosity at a given halo mass exhibits a large scatter, and the correlation between the two does not look very tight. Therefore, we also focus on the median of the host halo mass as a function of the luminosity, as depicted by the solid line.
At a high redshift ($z=2$), the dependence is rather weak. 
As the color indicates, the AGNs reside mainly in haloes with mass $\sim 10^{11}-10^{12.5} h^{-1}\,{\rm M}_{\odot}$.
At a relatively low redshift ($z=0.5$), the dependence of the median mass is somewhat stronger than that at $z=2$, with a nonmonotonic trend observed near the high luminosity end.
Beyond the median curve, there is a significant fraction of moderate and low luminosity AGNs with $40 \la \log (L_{\rm X} / \mathrm{erg~s^{-1}}) \la 43$ found in high mass haloes with $M_{\mathrm{halo}} \ga 10^{13} {\rm M}_{\odot}$, and the AGN host haloes have a broader distribution.
This trend in the dependence of the host halo mass on the AGN luminosity and redshift is qualitatively consistent with that of \citet{Gatti2016}, which have explored the AGN host halo mass with their semi-analytic model including AGN activities driven by galaxy interactions and disc instabilities.
The sharp cutoff of the distribution at the massive end of the halo mass in the top-right panel of Fig.~\ref{fig:lx_mhalo_4panel} corresponds to the fact that no haloes are found beyond this mass in our simulation at this redshift. Notably, low-luminosity AGNs can be found all the way until this limit.

In order to investigate the cause of the AGNs to be associated with haloes more massive than $M_{\mathrm{halo}} \sim 10^{13} {\rm M}_{\odot}$, we also show the $L_{\rm X}-M_{\mathrm{halo}}$ relation measured from galmodel in the bottom panels of Fig.~\ref{fig:lx_mhalo_4panel}.
In galmodel, in which we do not take into account the term $t_{\mathrm{loss}}$ in the gas accretion time-scale, the total number density of AGNs is lower than that in our fiducial model.
This is simply because the lifetime of the AGNs in galmodel is shorter than that in our fiducial model due to the absence of $t_{\mathrm{loss}}$.
The distribution of AGNs from galmodel looks different from that from the fiducial model in two ways. First, the distribution of the halo mass at a given luminosity decays more rapidly in galmodel towards the high mass end.
This indicates that the AGNs in high mass haloes as seen in the fiducial model have long gas accretion time-scales, and the effect of $t_{\mathrm{loss}}$ is significant at low redshift.
Second, the median curve of the host halo mass behaves differently in the two models. The median mass in the fiducial model is lifted from that in galmodel, again, because of the introduction of $t_\mathrm{loss}$ to allow massive haloes to populate less luminous AGNs.
At $z=2$, the impact of $t_{\mathrm{loss}}$ on the median is still moderate.
This is, however, much clearer at $z=0.5$ with the median increasing rapidly until $\log (L_{\rm X} / \mathrm{erg~s^{-1}}) \sim 42.5$, drops until $\log (L_{\rm X} / \mathrm{erg~s^{-1}}) \sim 44$, and flattens.
This trend is a unique feature only seen in the fiducial model.
The relatively faint AGNs in massive haloes with $M_{\mathrm{halo}} \ga 10^{13} {\rm M}_{\odot}$ are an important characteristic of the fiducial model which has a new time-scale, $t_{\mathrm{loss}}$.

\begin{figure*}
    \begin{center}
      \includegraphics[width=170mm]{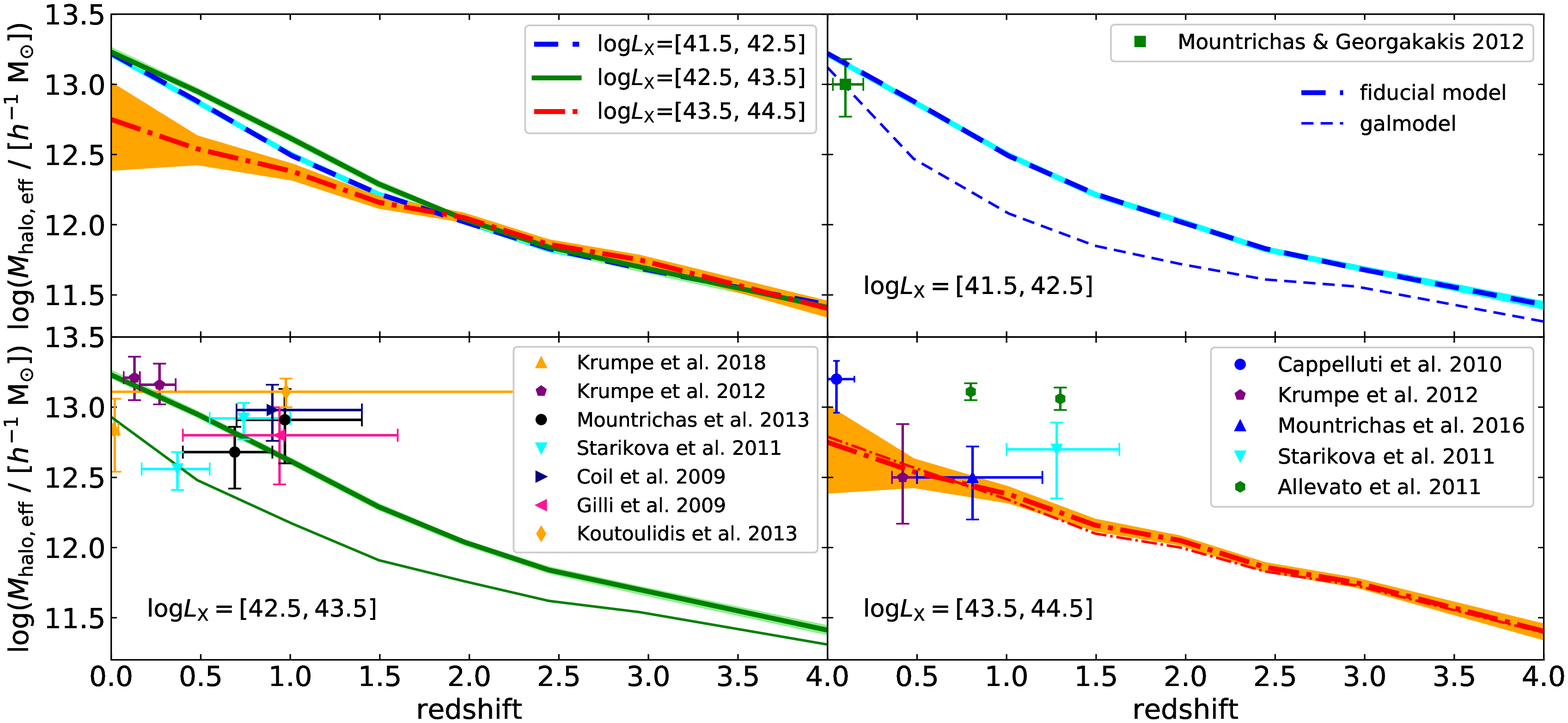}
    \end{center}
  \caption{Redshift evolution of the effective halo mass of AGNs.
  The lines in the top left panel depict those predicted by our fiducial model for AGNs with three luminosity ranges, $\log (L_{\rm X} / \mathrm{erg~s^{-1}}) =$ [41.5, 42.5] (blue dashed), [42.5, 43.5] (green solid), and [43.5, 44.5] (red dot-dashed).
  The shaded regions are the uncertainty of $M_{\mathrm{halo,eff}}$, which comes from the uncertainty on the AGN bias.
  The result is divided into the three luminosity ranges, top right, bottom left, and bottom right, respectively.
  The thin lines depict the result of galmodel.
  The filled points with error bars are the observational results, which are also divided into the three luminosity ranges based on the average 2-10~keV X-ray luminosity of the sample.
  The data is compiled by \citet{Georgakakis2019}.}
  \label{fig:z_meff_4panel}
\end{figure*}

\begin{figure*}
    \begin{center}
      \includegraphics[width=170mm]{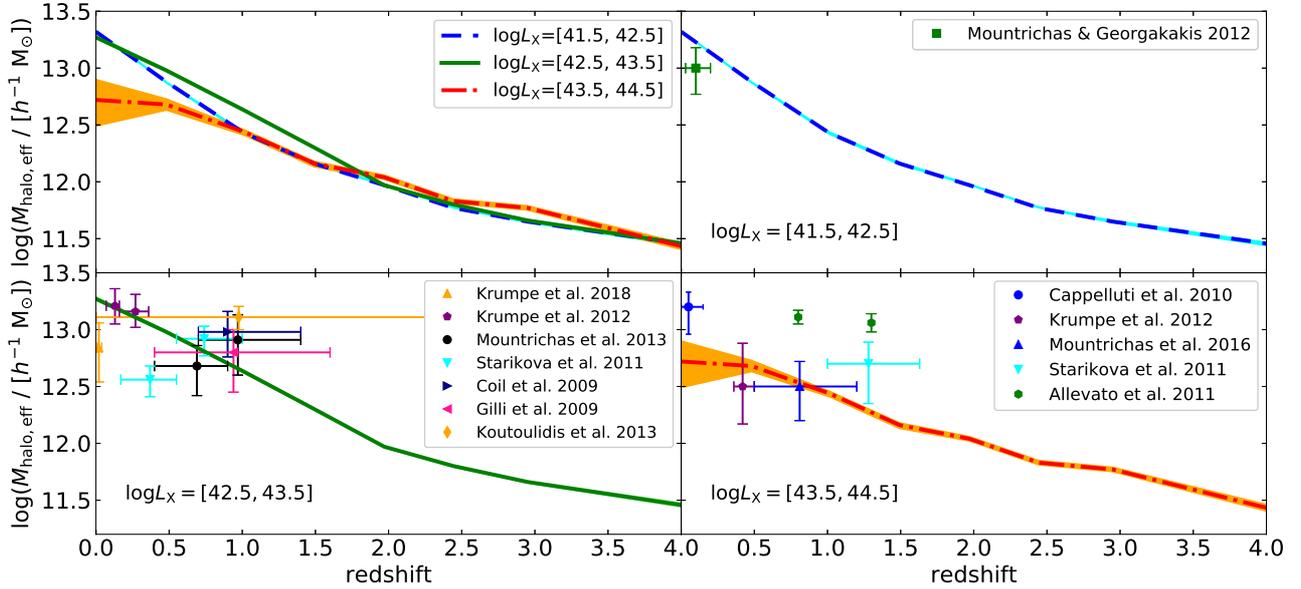}
    \end{center}
  \caption{Same as Fig.~\ref{fig:z_meff_4panel}, but for the effective halo mass derived from the power law fit to the 2PCFs.
  Only the results of the fiducial model are shown in this figure.
  }
  \label{fig:z_meff_from_power_law_fit_4panel}
\end{figure*}

\subsection{Effective halo mass}
\label{subsec:m_eff}
The wide distribution of the host halo mass given luminosity suggests that a single number such as
a typical halo mass alone cannot fully characterise the halo occupation properties of AGNs. Nevertheless, the effective halo mass is a quantity widely discussed in the literature and is indeed determined observationally in many previous analyses. Therefore, to see the consistency with the previous results, we here discuss this quantity in this subsection.

\subsubsection{Redshift dependence}

\begingroup
\renewcommand{\arraystretch}{1.3}
\begin{table*}
\caption{Results of power-law fits to the 2PCFs of X-ray AGNs of our model in 1120~$h^{-1}$~Mpc box.
}
\begin{center}
\begin{tabular}{lccccc}\hline\hline
Sample  & redshift & $\gamma$  & $r_{0}~(h^{-1}\,\mathrm{Mpc})$ & $b$ & $\log(M_{\mathrm{halo,eff}}/h^{-1}\,{\rm M}_{\odot})$ \\ \hline
$\log (L_{\rm X} / \mathrm{erg~s^{-1}}) = [41.5,42.5]$ & & \\
& 0.00 & $1.84^{+0.00}_{-0.00}$ & $5.95^{+0.02}_{-0.03}$ & $1.27^{+0.00}_{-0.01}$ & $13.32^{+0.01}_{-0.00}$ \\
& 0.48 & $1.77^{+0.00}_{-0.00}$ & $5.03^{+0.02}_{-0.02}$ & $1.38^{+0.00}_{-0.01}$ & $12.88^{+0.00}_{-0.01}$ \\
& 1.01 & $1.75^{+0.00}_{-0.00}$ & $4.31^{+0.02}_{-0.02}$ & $1.55^{+0.00}_{-0.00}$ & $12.43^{+0.01}_{-0.01}$ \\
& 1.49 & $1.72^{+0.00}_{-0.00}$ & $4.08^{+0.01}_{-0.02}$ & $1.79^{+0.01}_{-0.01}$ & $12.16^{+0.01}_{-0.01}$ \\
& 1.97 & $1.73^{+0.00}_{-0.00}$ & $4.05^{+0.01}_{-0.02}$ & $2.10^{+0.01}_{-0.01}$ & $11.97^{+0.01}_{-0.00}$ \\
& 2.44 & $1.77^{+0.00}_{-0.00}$ & $3.98^{+0.01}_{-0.02}$ & $2.39^{+0.01}_{-0.01}$ & $11.77^{+0.01}_{-0.00}$ \\
& 2.95 & $1.79^{+0.00}_{-0.00}$ & $4.11^{+0.02}_{-0.02}$ & $2.81^{+0.01}_{-0.01}$ & $11.65^{+0.01}_{-0.01}$ \\
& 4.04 & $1.91^{+0.01}_{-0.01}$ & $4.46^{+0.03}_{-0.04}$ & $3.92^{+0.03}_{-0.04}$ & $11.45^{+0.01}_{-0.02}$ \\ \hline

$\log (L_{\rm X} / \mathrm{erg~s^{-1}}) = [42.5,43.5]$ & & \\
& 0.00 & $1.81^{+0.01}_{-0.00}$ & $5.82^{+0.04}_{-0.04}$ & $1.24^{+0.01}_{-0.01}$ & $13.27^{+0.02}_{-0.01}$ \\
& 0.48 & $1.82^{+0.00}_{-0.01}$ & $5.30^{+0.03}_{-0.03}$ & $1.46^{+0.01}_{-0.01}$ & $12.98^{+0.01}_{-0.01}$ \\
& 1.01 & $1.81^{+0.00}_{-0.00}$ & $4.80^{+0.03}_{-0.03}$ & $1.72^{+0.01}_{-0.01}$ & $12.63^{+0.01}_{-0.01}$ \\
& 1.49 & $1.80^{+0.00}_{-0.00}$ & $4.40^{+0.02}_{-0.02}$ & $1.92^{+0.01}_{-0.01}$ & $12.30^{+0.00}_{-0.01}$ \\
& 1.97 & $1.75^{+0.00}_{-0.00}$ & $4.04^{+0.02}_{-0.02}$ & $2.10^{+0.01}_{-0.01}$ & $11.97^{+0.01}_{-0.01}$ \\
& 2.44 & $1.77^{+0.01}_{-0.00}$ & $4.05^{+0.02}_{-0.02}$ & $2.42^{+0.01}_{-0.01}$ & $11.80^{+0.01}_{-0.01}$ \\
& 2.95 & $1.81^{+0.00}_{-0.00}$ & $4.14^{+0.03}_{-0.03}$ & $2.84^{+0.02}_{-0.02}$ & $11.66^{+0.01}_{-0.01}$ \\
& 4.04 & $1.92^{+0.01}_{-0.01}$ & $4.46^{+0.04}_{-0.04}$ & $3.92^{+0.03}_{-0.04}$ & $11.45^{+0.02}_{-0.02}$ \\

\hline

$\log (L_{\rm X} / \mathrm{erg~s^{-1}}) = [43.5,44.5]$ & & \\
& 0.00 & $1.81^{+0.08}_{-0.08}$ & $4.49^{+0.27}_{-0.33}$ & $0.98^{+0.07}_{-0.07}$ & $12.72^{+0.18}_{-0.23}$ \\
& 0.48 & $1.93^{+0.02}_{-0.02}$ & $4.41^{+0.08}_{-0.10}$ & $1.26^{+0.03}_{-0.03}$ & $12.68^{+0.05}_{-0.05}$ \\
& 1.01 & $1.89^{+0.01}_{-0.02}$ & $4.26^{+0.04}_{-0.05}$ & $1.56^{+0.02}_{-0.02}$ & $12.44^{+0.02}_{-0.03}$ \\
& 1.49 & $1.86^{+0.01}_{-0.01}$ & $4.03^{+0.05}_{-0.05}$ & $1.79^{+0.02}_{-0.02}$ & $12.16^{+0.02}_{-0.03}$ \\
& 1.97 & $1.86^{+0.01}_{-0.01}$ & $4.17^{+0.04}_{-0.05}$ & $2.18^{+0.02}_{-0.03}$ & $12.04^{+0.02}_{-0.02}$ \\
& 2.44 & $1.81^{+0.01}_{-0.01}$ & $4.11^{+0.04}_{-0.05}$ & $2.46^{+0.03}_{-0.03}$ & $11.83^{+0.02}_{-0.02}$ \\
& 2.95 & $1.86^{+0.01}_{-0.01}$ & $4.39^{+0.04}_{-0.04}$ & $3.01^{+0.03}_{-0.03}$ & $11.77^{+0.02}_{-0.02}$ \\
& 4.04 & $1.89^{+0.01}_{-0.01}$ & $4.40^{+0.06}_{-0.07}$ & $3.86^{+0.06}_{-0.06}$ & $11.42^{+0.03}_{-0.03}$ \\

\hline
\label{tbl:fitting_results}
\end{tabular}
\end{center}
\end{table*}
\endgroup

We show the redshift and luminosity dependence of the effective halo mass, $M_{\mathrm{halo,eff}}$, in the top left panel of Fig.~\ref{fig:z_meff_4panel} for three different luminosity bins as indicated in the figure legend.
The shaded regions show the error bar in $M_{\mathrm{halo,eff}}$, which originates from that in the AGN bias averaged from 1 to 20~$h^{-1}$~Mpc, although it is very small and difficult to see for the moderate and low luminosity bins due to the large sample size.
At first glance $M_{\mathrm{halo,eff}}$ increases with cosmic time in all luminosity ranges.
From $z\sim 4$ to $z\sim 2$, $M_{\mathrm{halo,eff}}$ shows no significant luminosity dependence, and increases from $\sim 10^{11.3} h^{-1}\,{\rm M}_{\odot}$ to $\sim 10^{12} h^{-1}\,{\rm M}_{\odot}$.
At redshift lower than $z\sim 2$, the shaded regions for the different luminosity bins start to deviate from each other; the low luminosity bin shows larger host halo masses than the most luminous bin ($\log (L_{\rm X} / \mathrm{erg~s^{-1}}) \geq 43.5$).
We observe here a negative dependence of the AGN bias on the AGN luminosity.
While $M_{\mathrm{halo,eff}}$ is at most $\sim 10^{12.7} h^{-1}\,{\rm M}_{\odot}$ for the highest luminosity bin, the other two bins can reach $\sim 10^{13.25} h^{-1}\,{\rm M}_{\odot}$ at $z=0$.
The high $M_{\mathrm{halo,eff}}$ for the AGNs in the moderate luminosity bin is expected from the
increase of the median halo mass of the AGN with $\log (L_{\rm X} / \mathrm{erg~s^{-1}}) \sim 42.5$ as shown in the top right panel of Fig.~\ref{fig:lx_mhalo_4panel}.
For the low luminosity AGNs, the median halo mass is smaller than or comparable to that of the luminous AGNs, nevertheless the effective halo mass is larger.
As will be discussed in Section 3.4, this is partly because of the difference between the mean and the median for a skewed distribution. In addition, the effective halo mass is derived based on the number-weighted mean of the bias, not the mass. The non-linear dependence of the bias on the halo mass could add a complication to the interpretation of the effective mass derived this way.
In Section 3.4, we will argue the effect of the skewness of the host halo mass distributions of AGNs on the effective halo mass.
The high $M_{\mathrm{halo,eff}}$ for moderate and low luminosity AGNs at low redshift is consistent with the host halo mass observationally inferred (i.e., $\sim10^{13} h^{-1}\,{\rm M}_{\odot}$) as described in Section~1.

We compare $M_{\mathrm{halo,eff}}$ obtained with our fiducial model (thick lines) to that with galmodel (thin lines) in the rest of the panels of Fig.~\ref{fig:z_meff_4panel}, showing the results with three luminosity ranges now in a panel for each.
While there is not significant difference for the highest luminosity range, $M_{\mathrm{halo,eff}}$ from our fiducial model is larger than that from galmodel for the lowest and intermediate luminosity ranges, in particular, at low redshift.
For the latter two luminosity ranges, it is $\sim~0.4$~dex  ($\sim~0.2$~dex) larger than galmodel at $z \la 2$ ($z \ga 2$).

We then compare the model predictions with observations in the top-right, bottom-left, and bottom-right panels of Fig.~\ref{fig:z_meff_4panel}.
The filled circles with error bars show the observational data, which we divide into the three luminosity bins based on the median value from each observation.
The observational data is taken from \citet{Georgakakis2019}, who have compiled the data in the literature (\citealt{Gilli2009}; \citealt{Coil2009}; \citealt{Cappelluti2010}; \citealt{Allevato2011}; \citealt{Starikova2011}; \citealt{Krumpe2012}; \citealt{Mountrichas2012}; \citealt{Koutoulidis2013}; \citealt{Mountrichas2013}; \citealt{Mountrichas2016}; \citealt{Krumpe2018}).
Overall, our model is consistent with the current measurements at $z \la 1.5$, that is, the typical host halo mass of $10^{12.5-13.5} h^{-1}\,{\rm M}_{\odot}$.
The agreement between the prediction and the observations is clearly better for the fiducial model than galmodel in the lower left panel for the intermediate luminosity bin.
Further observations with more accuracy and up to a higher redshift especially for the lowest luminosity bin are expected to provide key information to constrain these theoretical models.

For more direct comparison with the observations, we show $M_{\mathrm{halo,eff}}$ derived from the AGN bias based on our power-law fits to $w_p (r_p)$ in Fig.~\ref{fig:z_meff_from_power_law_fit_4panel}.
In Table \ref{tbl:fitting_results} we list the redshift, the derived $\gamma$, $r_0$, $b$ from Eqn. \ref{eq:bias_power_law}, and $M_{\mathrm{halo,eff}}$ inferred from the derived best-fitting value of $b$.
Through this procedure, we can also examine possible systematic effects originating from
the power-law fit on the estimation of the AGN bias and the effective halo mass.
For AGNs in all the luminosity ranges, the estimated $M_{\mathrm{halo,eff}}$ is similar to that derived from the AGN bias averaged from 1 to 20~$h^{-1}$~Mpc (Fig.~\ref{fig:z_meff_4panel}).
The shaded regions in Fig.~\ref{fig:z_meff_from_power_law_fit_4panel} show the 1 $\sigma$ error of $M_{\mathrm{halo,eff}}$, which is propagated from those of $\gamma$ and $r_0$.

\subsubsection{Luminosity dependence}

\begin{figure*}
    \begin{center}
      \includegraphics[width=170mm]{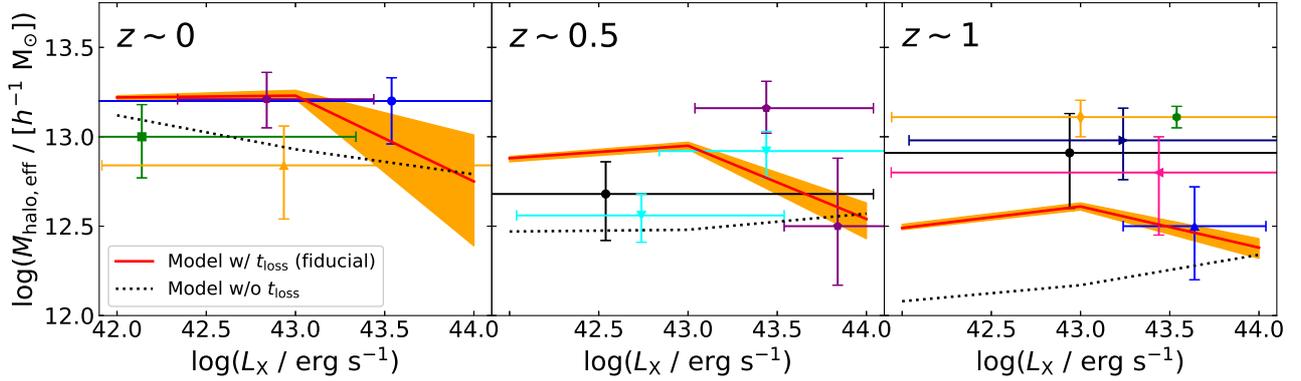}
    \end{center}
  \caption{The effective halo mass of AGN, $M_{\mathrm{halo,eff}}$, as a function of 2-10~keV X-ray luminosity.
  The red solid lines depict those predicted by our fiducial model.
  The shaded regions are the uncertainty of $M_{\mathrm{halo,eff}}$, which comes from the uncertainty on the AGN bias.
  The black dotted lines show the result of galmodel.
  The results at three redshifts are plotted; $z=0$ (left), $0.5$ (middle), and $1$ (right), respectively.
  Various colored points with error bars are the observational results, which are also divided into the three redshift ranges based on the median redshift of the sample.
  The colours and symbols are the same as in Fig.~\ref{fig:z_meff_4panel} and \ref{fig:z_meff_from_power_law_fit_4panel}.
  The data is compiled by \citet{Georgakakis2019}.
  The error bars in the horizontal direction indicate the range of luminosities in each sample, most of which is taken from the data complied by \citet{Fanidakis2013a} (their Table 1).
  That of \citet{Mountrichas2016} is taken from their original paper.}
  \label{fig:lumx_meff_3panel}
\end{figure*}

\begin{figure*}
    \begin{center}
      \includegraphics[width=170mm]{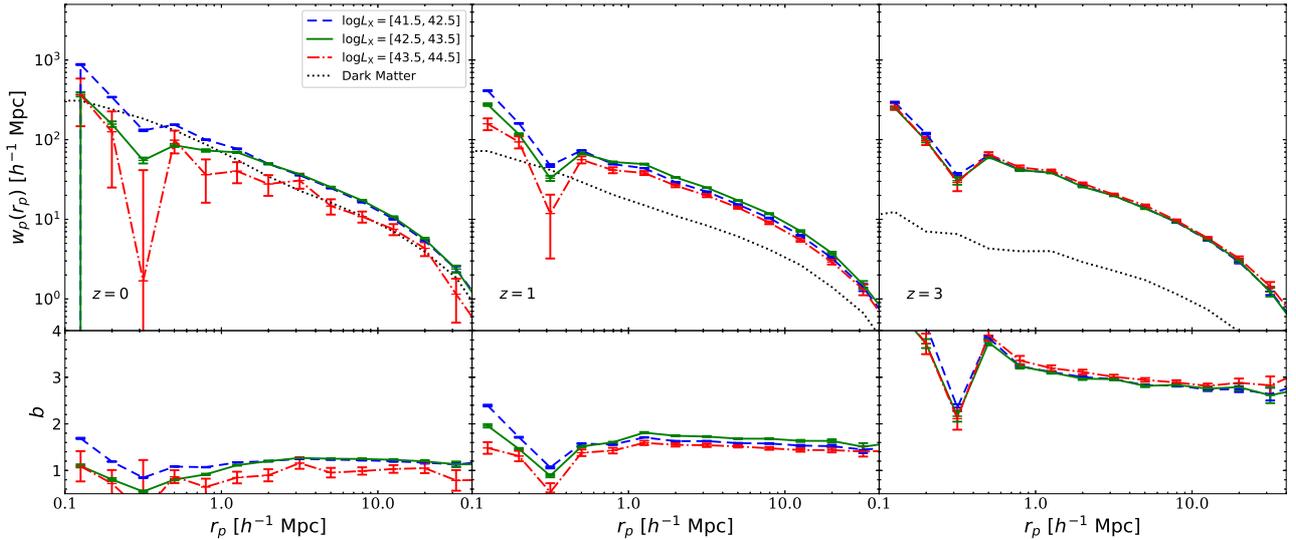}
    \end{center}
  \caption{Upper panels: from left to right, two-point correlation functions of AGNs predicted by our model at $z=0$, 1, and 3.
  The blue, green, and red lines depict those of AGNs with luminosity ranges, $\log (L_{\rm X} / \mathrm{erg~s^{-1}}) =$ [41.5, 42.5], [42.5, 43.5], and [43.5, 44.5], respectively.
  The error bars are calculated by jackknife
  resampling of eight sub-volumes.
  The dotted lines are the two-point correlation function of dark matter particles of the $\nu^2$GC simulation.
  Lower panels: the bias of the three AGN samples as a function of separation $r_p$.}
  \label{fig:2pcf_mod}
\end{figure*}

\begin{figure*}
    \begin{center}
      \includegraphics[width=170mm]{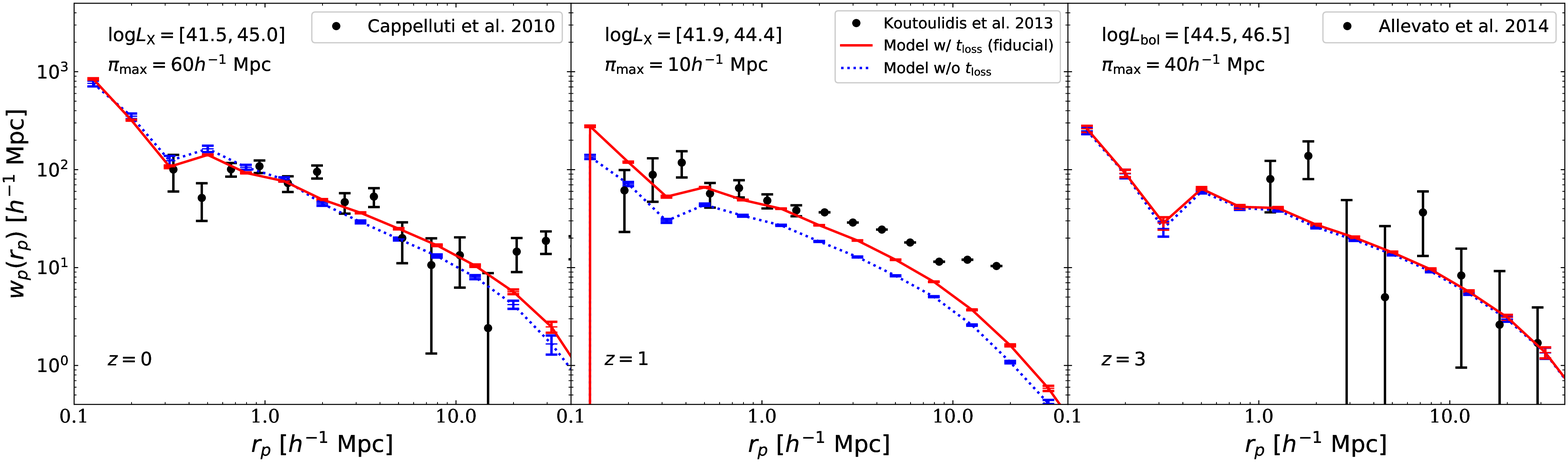}
    \end{center}
  \caption{From left to right, two-point correlation functions of AGNs at $z=0$, 1, and 3.
  The red solid and blue dotted lines show the results of the fiducial model and galmodel.
  The error bars are calculated by jackknife resampling of eight sub-volumes.
  The filled circles with error bars are the observational result
  from \citet{Cappelluti2010} ($z \sim 0$), \citet{Koutoulidis2013} ($z \sim 1$), and \citet{Allevato2014} ($z \sim 3$).
  The luminosity range and $\pi_{\mathrm{max}}$ in Eqn.~\ref{eq:wp} are adopted to match each observation.}
  \label{fig:2pcf_obs}
\end{figure*}

We now discuss the luminosity dependence of $M_{\mathrm{halo,eff}}$, comparing our model results with observations.
We plot the observational data as a function of $L_{\rm X}$ in three redshift ranges ($z=0-0.25$, $0.25-0.75$ and $0.75-1.25$) in Fig.~\ref{fig:lumx_meff_3panel}. 
Each of the samples is assigned to one of the three redshift bins based on the median redshift.
We also calculate $M_{\mathrm{halo,eff}}$ in our model at $z=0$, $0.5$, and $1.0$, and  plot them by the solid (fiducial, with $1$-$\sigma$ uncertainty by the shade) and the dotted (galmodel) lines.
Overall, the results of our fiducial model are consistent with observations.
In particular, our model predicts a negative dependence of $M_{\mathrm{halo,eff}}$ on the AGN luminosity at low redshift: in all the panels, $M_{\mathrm{halo,eff}}$ of AGNs with $\log (L_{\rm X} / \mathrm{erg~s^{-1}}) \sim 43$ is higher than that of more luminous AGNs with $\log (L_{\rm X} / \mathrm{erg~s^{-1}}) \sim 44$.
This result reflects the existence of less luminous AGNs in massive haloes with $M_{\mathrm{halo}} \ga 10^{13} {\rm M}_{\odot}$, as shown in Section \ref{subsec:lx_mhalo}.
This luminosity dependence is qualitatively in agreement with the observational result of \citet{Mountrichas2016} and the model of \citet{Fanidakis2013a}.
In contrast, in galmodel, in which we do not take into account the term $t_{\mathrm{loss}}$ in the gas accretion time-scale, $M_{\mathrm{halo,eff}}$ positively depends on AGN luminosity ($z=0.5$ and $1$) or do not have significant luminosity dependence ($z=0$).
This result also reflects the deficiency of less luminous AGNs in massive haloes (the bottom right panel of Fig.~\ref{fig:lx_mhalo_4panel}).
The fiducial model is clearly a better representation of the data at $z\sim1$.
At lower redshifts, our fiducial model, however, overpredicts the observed effective halo mass of AGNs at low luminosity ($\log (L_{\rm X} / \mathrm{erg~s^{-1}}) < 43$) compared to galmodel.
We note that the observational data shown here is based on AGN samples in a wide redshift range to reduce the statistical error of $M_{\mathrm{halo,eff}}$.
Thus, the validity of the predicted negative luminosity dependence of $M_{\mathrm{halo,eff}}$ 
is still inconclusive observationally.
Therefore a more accurate determination of the luminosity dependence of $M_{\mathrm{halo,eff}}$ is expected to provide key information to constrain theoretical models.

\subsection{Two-point correlation functions of X-ray AGNs}
\label{subsec:wp_xray}

In this subsection, we discuss the 2PCFs of AGNs.
Our high resolution cosmological $N$-body simulation covering covering a cubic gigaparsec comoving volume enables us to analyse the clustering of AGNs, which are rare objects, with the AGN auto-correlation function.
In the top panels of Fig.~\ref{fig:2pcf_mod}, we plot the 2PCF of AGN divided into three X-ray luminosity bins and 2PCF of DM particles measured from the $\nu^2$GC simulation.
At low redshifts ($z=0, 1$), the clustering amplitude of less luminous AGNs is slightly higher than luminous AGNs.
On the other hand, at $z\sim 3$, the 2PCF is almost independent of the AGN luminosity, which is consistent with no significant luminosity dependence of $M_{\mathrm{halo,eff}}$ as shown in Fig.~\ref{fig:z_meff_4panel}.
In the bottom panels of Fig.~\ref{fig:2pcf_mod}, the AGN bias, $(w_{p, \mathrm{AGN}} (r_p) / w_{p, \mathrm{AGN}} (r_p))^{1/2}$, is plotted.
While the dependence of the bias on the luminosity is observed to be weak, there appears clear increase with increasing redshift. 
In all the panel, there is a sudden upturn in of the bias towards the left end of the plots, below several hundred $h^{-1}$~kpc. This might be a hint of the transition between one and two halo regime.

Now, let us discuss if these predictions make sense by comparing with observations.
We compare the 2PCFs from our model with observations directly.
We focus on the auto-correlation functions of AGNs measured in \citet{Cappelluti2010}, \citet{Koutoulidis2013}, and \citet{Allevato2014} ($z \sim 0, 1, 3$) instead of the AGN-galaxy cross-correlation functions.
For a fair comparison, we adopt the same ranges of luminosity and $\pi_{\mathrm{max}}$ in Eqn.~\ref{eq:wp} as in the corresponding observation.
Because \citet{Koutoulidis2013} have used a sample selected in the 0.5-8keV band, we convert $L_{\rm X}(0.5-8~\mathrm{keV})$ to $L_{\rm X}(2-10~\mathrm{keV})$ assuming the photon index $\Gamma = 1.4$.
We here show that the results of our fiducial model (red solid lines) are in good agreement with observed AGN clustering in Fig.~\ref{fig:2pcf_obs}.
In this figure, the results from galmodel are also plotted (blue dotted lines).
At $z=0$, there is a significant difference between the two models, although the current observation \citep{Cappelluti2010} cannot tell which fits better due to the large statistical error.
At $z=1$, galmodel clearly underestimate the amplitude of $w_p (r_p)$, which is inconsistent with the result of \citet{Koutoulidis2013}.
While the 2PCF on small scale ($\la 1\, h^{-1}$~Mpc) supports our fiducial model over galmodel, the amplitude on larger scales is still lower compared with the observation.
This may be partly due to the wide redshift range of the sample of \citet{Koutoulidis2013}, which is shown by the horizontal error bar in the bottom left panels of Fig.~\ref{fig:z_meff_4panel} and \ref{fig:z_meff_from_power_law_fit_4panel}.
Actually, in Fig.~\ref{fig:z_meff_4panel}, the predicted $M_{\mathrm{halo,eff}}$ at $z\sim0$ is consistent with their result. 
In observations, one analyses AGNs distributed over finite
redshift ranges, and consequently the 2PCF derived from such a sample is a composite of 2PCFs at different redshifts.
On the other hand, in simulations we use an AGN sample at an exact redshift to represent the observation, although how good a representative such a mock sample is must be carefully investigated, especially when the redshift range is wide.
We need to construct a more realistic mock AGN catalog which takes into account the light-cone effect for the complete comparison between the model and the observation.
We leave this issue as a future work.
At $z=3$, both models are within the large uncertainty of the observation \citep{Allevato2014}.
Further studies of 2PCF with larger AGN samples, including the luminosity and redshift dependence, can constrain the theoretical models.

\subsection{AGN host halo mass distribution}
\label{subsec:halo_mass_dist}

\begin{figure*}
    \begin{center}
      \includegraphics[width=170mm]{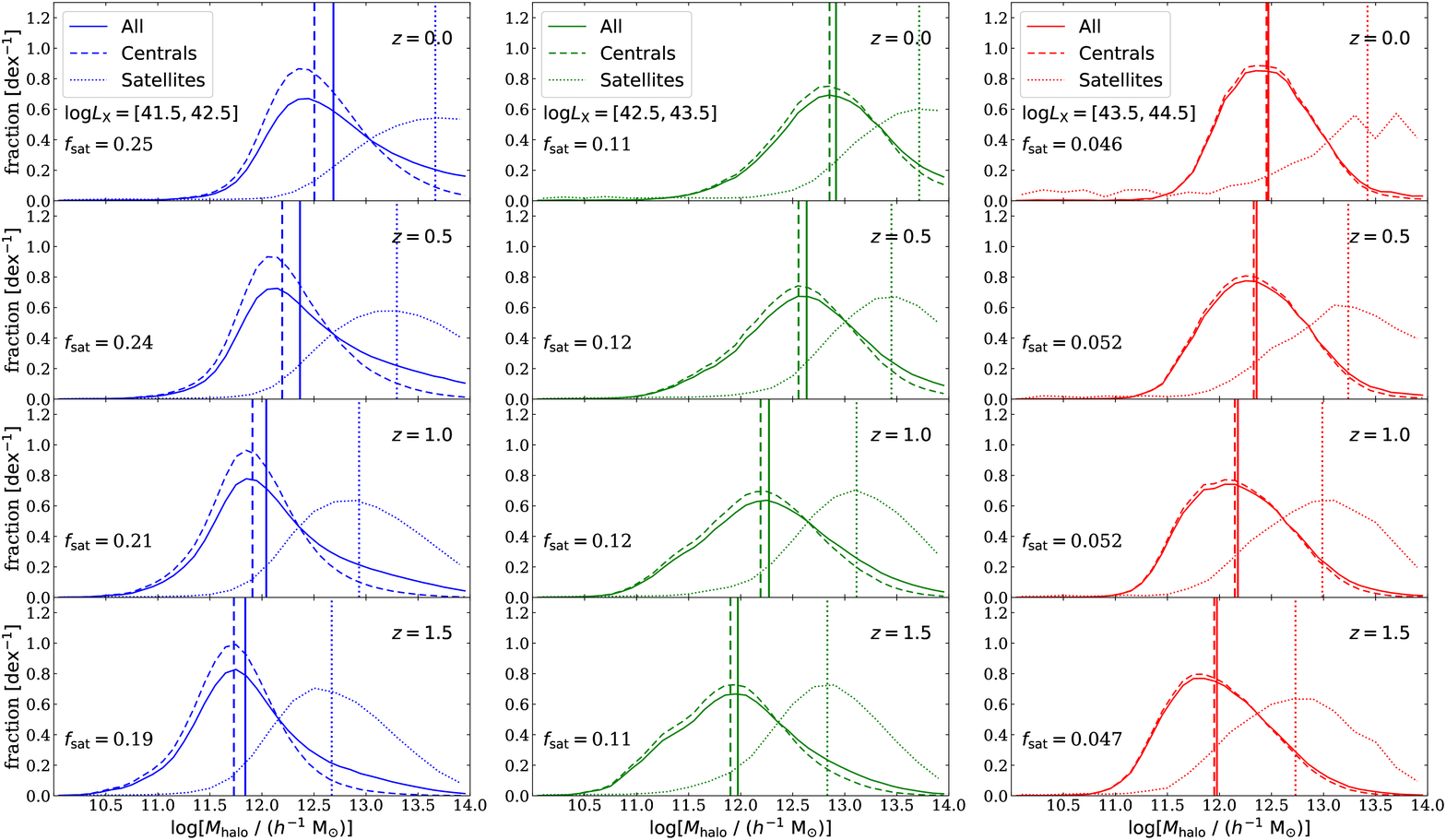}
    \end{center}
  \caption{Host halo mass distributions of AGNs with X-ray luminosity ranges of $\log (L_{\rm X} / \mathrm{erg~s^{-1}}) =$ [41.5, 42.5] (left), [42.5, 43.5] (centre), and [43.5, 44.5] (right) at at $z=0$, $0.5$, $1.0$, and $1.5$ (from top to bottom).
  The solid, dashed, and dotted lines depict those of all, central, and satellite AGNs, respectively.
  The vertical line shows the median halo mass for each distribution.}
  \label{fig:m_halo_dist}
\end{figure*}

In this subsection, we explore the mass distribution of the DM haloes hosting AGNs with given luminosities and its redshift evolution.
In contrast to observational studies, we can directly derive these because our model is based on the merger trees of DM haloes.
We can also compare the median halo mass of the distribution with the effective halo mass $M_{\mathrm{halo,eff}}$ which we have derived in Section 3.2.
This allows us to check the validity of the method to derive the typical host halo mass of AGNs from the clustering bias, which is frequently employed in observational studies the literature.
We show the host halo mass distributions of our model AGNs in Fig.~\ref{fig:m_halo_dist} (solid line).
This figure shows that the host halo mass distributions, when plotted as a function of the logarithm of the host halo mass, are skewed and have a significantly heavy tail towards higher mass, in particular, for low luminosity AGNs and at low redshifts.
We show the median halo masses of AGNs by the vertical solid lines. Due to the positively skewed distribution, the locations of the median are significantly larger than those of the peaks.
The median halo masses of AGNs for all the three luminosity ranges increase with cosmic time.
For example, the median halo mass for low luminosity AGNs increases from $10^{11.8} h^{-1}\,{\rm M}_{\odot}$ at $z=1.5$ to $10^{12.7} h^{-1}\,{\rm M}_{\odot}$ at $z=0$.

We also separately show the host halo mass distributions of central (dashed) and satellite (dotted) AGNs.
The distributions are normalised to unity when integrated over the logarithm of the halo mass.
We also indicate the fraction of satellite AGNs to the total AGN population in the figure legend in each panel.
We discuss the satellite fraction as a function of redshift in Section \ref{subsubsec:sat_frac}.
Here, central AGNs are defined as those which live in a central galaxy, and satellite AGNs correspond to to those in satellite galaxies.
In a given X-ray luminosity range, satellite AGNs tend to reside in more massive haloes than central AGNs.
This is due to the following reasons.
The stellar mass distributions of the host galaxies of the satellite AGNs and central AGNs in the same X-ray luminosity range are similar in our model.
In this case, the host halo mass of a satellite AGN when it was a central before merging into a more massive galaxy is expected to be similar to those of the other central AGNs.
Thus, a halo hosting the satellite AGNs is typically more massive than that having a central AGNs with the same luminosity.
As a consequence, a typical host halo mass of the satellite AGNs is higher than that of the central AGNs.
The positive skewness of the host halo mass distribution is mainly due to the contribution from these satellite AGNs.
The median halo masses of satellite AGNs, marked by the vertical dotted lines, are $\sim$0.7-1.0~dex higher than those of central AGNs (the vertical dashed lines).
There is significant contribution of the halo mass distribution of satellite to the median of all sample for faint AGNs, while little contribution for bright AGNs.
We will investigate the origin of the satellite AGNs in Section 3.6.

\begin{figure}
	 \centering
	\includegraphics[width=0.9\columnwidth]{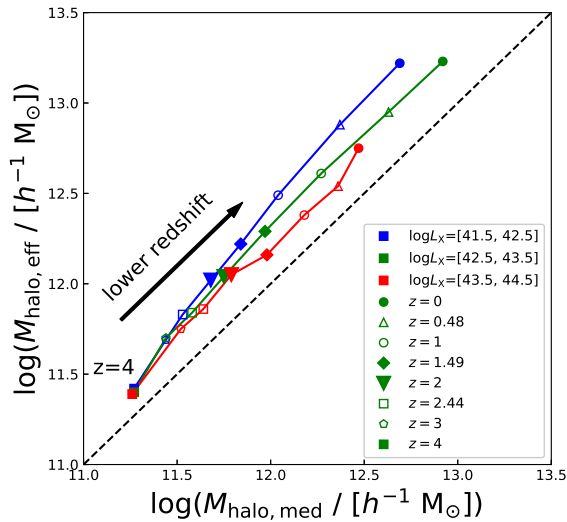}
    \caption{Relation between the median halo mass and the effective halo mass for AGNs with luminosity ranges, $\log (L_{\rm X} / \mathrm{erg~s^{-1}}) =$ [41.5, 42.5] (blue), [42.5, 43.5] (green), and [43.5, 44.5] (red) at various redshifts.
    }
    \label{fig:mmed_meff}
\end{figure}

In Fig.~\ref{fig:mmed_meff}, we show the relation between the median halo mass, $M_{\mathrm{halo,med}}$, and the effective halo mass, $M_{\mathrm{halo,eff}}$, of given luminosities from $z=4$ to $z=0$ in our model.
For all the cases, $M_{\mathrm{halo,eff}}$ is estimated to be higher than $M_{\mathrm{halo,med}}$.
Moreover, the difference is larger for lower redshift and lower luminosity.
The degree of increase in $M_{\mathrm{halo,eff}}$ as decreasing redshift is smaller than that in $M_{\mathrm{halo,med}}$, in particular, for low luminosity AGNs.
This result can be understood as a consequence of the larger fraction of satellites at low redshift for the fainter AGNs, which causes the skewness of the host halo mass distribution.
For example, while $M_{\mathrm{halo,med}}$ of low luminosity AGN at $z=1$ (blue open circle) is lower than that of luminous AGN (red open circle), $M_{\mathrm{halo,eff}}$ of the former is higher than that of the latter (compare the circles in different colours).
The skewness (and/or broadness) of the host halo mass distribution of the low luminosity AGN (third from the top of the left column of Fig.~\ref{fig:m_halo_dist}) may account for the high effective halo mass.
This analysis indicates that the effective halo mass is systematically higher than the median halo mass, and that the difference can reach $\sim$~0.5~dex, in particular, at lower redshift and for low luminosity AGNs.

As shown above, the median halo mass is significantly smaller than the effective halo mass derived from the AGN bias.
This result is consistent with \citet{Leauthaud2015}, who use weak lensing measurements and an abundance-matching approach to investigate the AGN host haloes.
Their technique can infer the full halo mass distribution for AGNs in contrast to most of previous studies, which have inferred a single effective halo mass from AGN clustering (for recent attempts to determine the HOD of AGNs from clustering, see also \citealt{Starikova2011}; \citealt{Richardson2012}; \citealt{Allevato2012}; \citealt{Chatterjee2013}; \citealt{Krumpe2018}).
They have focused on moderate luminosity X-ray AGNs at $z<1$ from the COSMOS field, and have found that the median halo mass is $\log (M_{\mathrm{halo}} / {\rm M}_{\odot}) \sim 12.5$, which is smaller than the effective halo mass of their sample, $\log (M_{\mathrm{halo}} / {\rm M}_{\odot}) \sim 12.7$.
They also have found that the effective halo mass is in between the median and the mean halo mass.
Here we confirmed their trend; the mean halo masses are roughly 0.3~dex (0.2~dex) larger than the effective halo masses at $z \la 2$ (z \ga 2.5) in all luminosity ranges considered in this paper ($\log (L_{\rm X} / \mathrm{erg~s^{-1}}) =$ [41.5, 44.5]).

\subsection{Halo occupation distribution of X-ray AGNs}
\label{subsec:hod_xray}

\begin{figure*}
    \begin{center}
      \includegraphics[width=170mm]{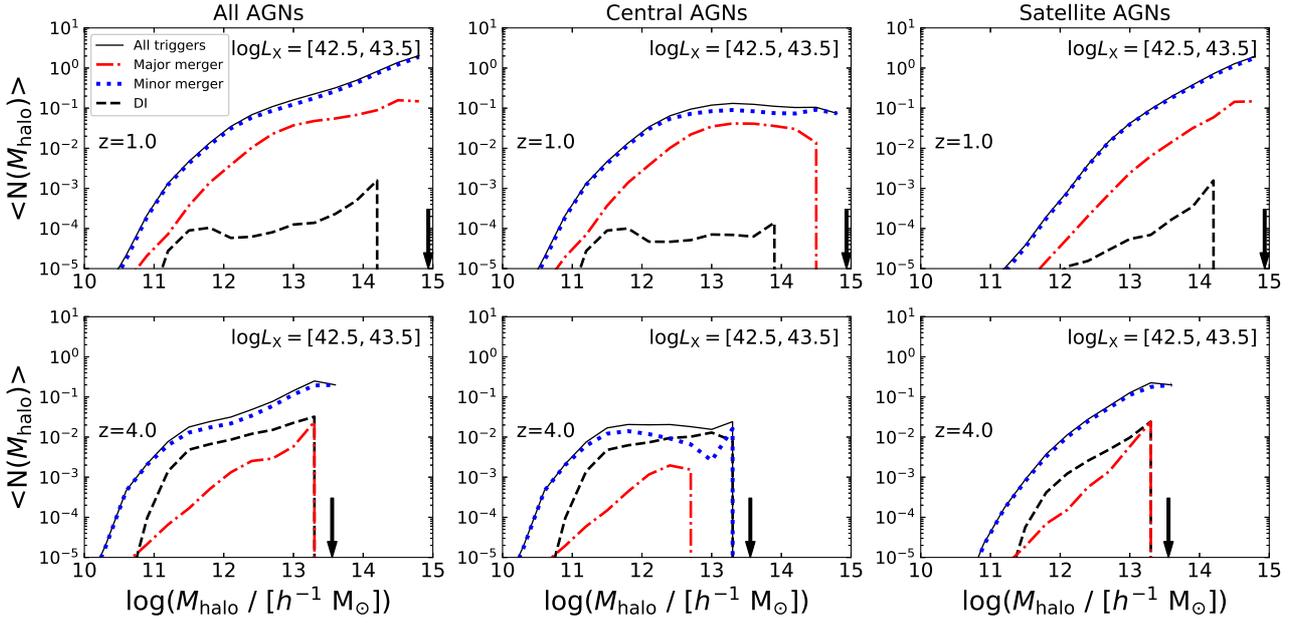}
    \end{center}
  \caption{Halo occupation distributions of our model AGNs with $\log (L_{\rm X} / \mathrm{erg~s^{-1}}) = [42.5, 43.5]$ at $z=1$ (top) and $4$ (bottom).
  The three panels in each row depict those of all (left), central (centre), and satellite (right) AGNs.
  The HODs are divided by the triggering mechanisms, major mergers (red dotted-dashed), minor mergers (blue dotted), disc instabilities (black dashed, shown as DI in the legend), and all triggering mechanisms (black thin solid).
  Black arrows show the maximum halo mass at these redshifts in the simulation.
  }
  \label{fig:hod_midlum_mod}
\end{figure*}

\begin{figure*}
    \begin{center}
      \includegraphics[width=170mm]{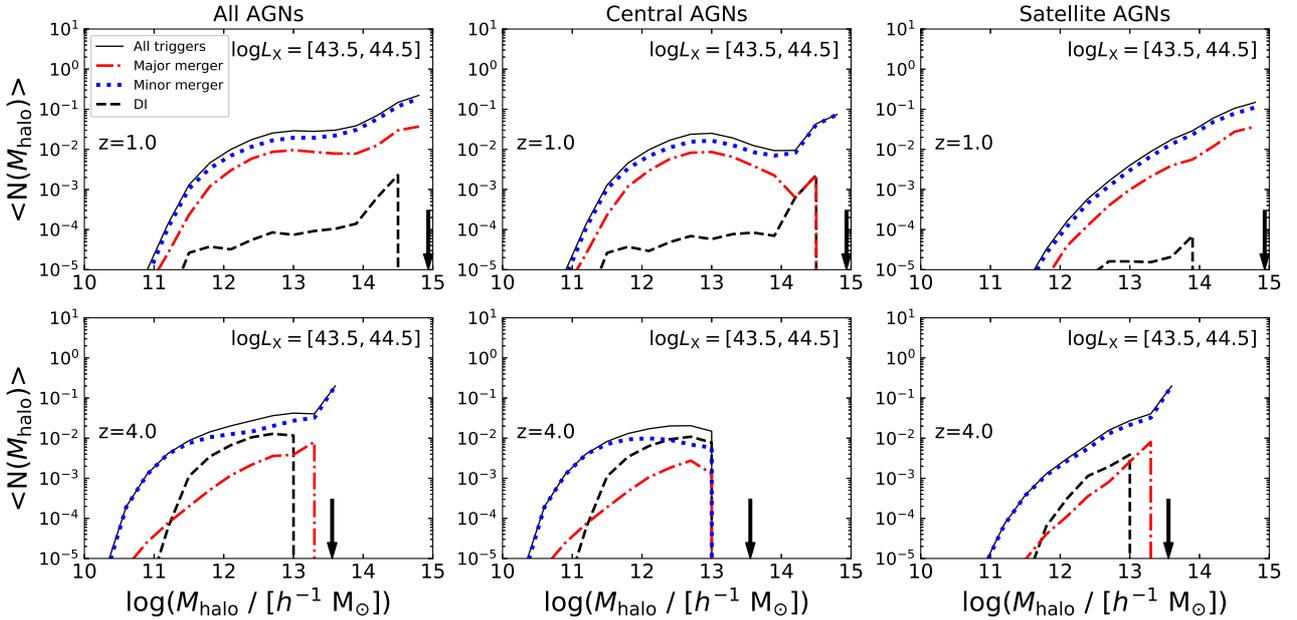}
    \end{center}
  \caption{Same as Fig.~\ref{fig:hod_midlum_mod}, but for the AGNs with $\log (L_{\rm X} / \mathrm{erg~s^{-1}}) = [43.5, 44.5]$.
  }
  \label{fig:hod_highlum_mod}
\end{figure*}

\begin{figure*}
    \begin{center}
      \includegraphics[width=170mm]{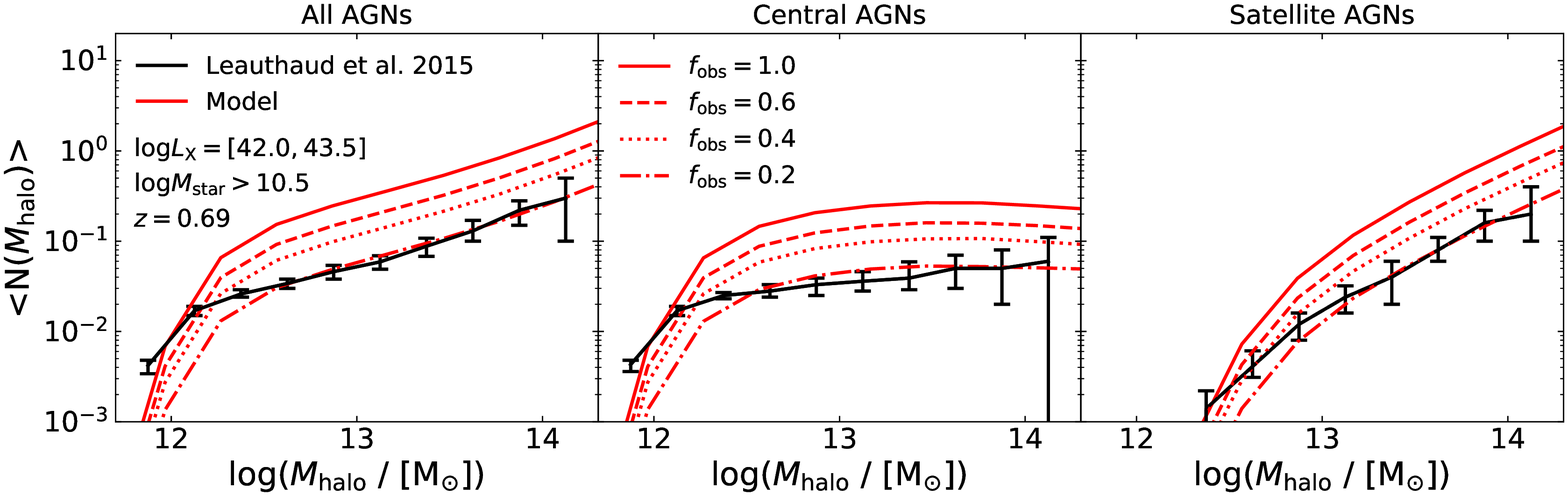}
    \end{center}
  \caption{The HODs predicted by our model, compared with those inferred from observations (Leauthaud et al. 2015).
  The three panels show the HODs of all (left), central (middle), and satellite (right) AGNs, respectively.
  Red lines depict those predicted by our model for varying obscured fractions from 0.2 (dot-dashed), 0.4 (dotted), 0.6 (dashed), and 1.0 (solid).
  Black solid lines with error bars are the result from \citet{Leauthaud2015}.
  }
  \label{fig:hod_comparison}
\end{figure*}

\subsubsection{Model prediction}
\label{subsubsec:model}

The HODs are expected to be one of the important constraints on the AGN formation models as well as the 2PCFs.
Because the HOD gives the fraction of active haloes (hosting AGNs) to the total number of haloes, it corresponds to the duty cycle of AGN activity.
Observationally, there is degeneracy in the functional forms of the HOD 
(\citealt{Kayo2012}; \citealt{Richardson2012}; \citealt{Shen2013}), in particular, the satellite HODs  such that different HODs can reproduce the 2PCF equally well.
Recent observational studies, however, have tried to obtain HODs from direct measurements (\citealt{Allevato2012}; \citealt{Chatterjee2013}) or from weak lensing measurements and an abundance matching technique (\citealt{Leauthaud2015}).
Future improvement on the HOD measurement could give us important clues to the understanding of the AGN formation.

Firstly, we show the predicted HODs with two luminosity bins, $\log (L_{\rm X} / \mathrm{erg~s^{-1}}) = [42.5, 43.5], \ [43.5, 44.5]$, at $z=1$ and $4$ as representatives in Fig.~\ref{fig:hod_midlum_mod} and \ref{fig:hod_highlum_mod} with no observational selection cuts, so as to capture general features of the HODs of our model.
The left panels of Fig.~\ref{fig:hod_midlum_mod} show the mean occupation number of AGNs, $\langle N \rangle$, with $\log (L_{\rm X} / \mathrm{erg~s^{-1}}) = [42.5, 43.5]$ at $z=1$ and 4.
We show four types of HODs for each panel: the HODs of AGNs triggered by major mergers (red dot-dashed), minor mergers (blue dotted), disc instabilities (black dashed), and all triggering mechanisms (thin black solid line).
These plots represent typical environments where the three mechanisms are efficient in triggering AGN activity.
In this luminosity range, the mean occupation number of central AGNs, $\langle N_{\mathrm{cen}} \rangle$ (thin black solid lines of the middle panels), increases with $M_{\mathrm{halo}}$, rises up to $\sim 10^{-1}$ ($\sim 10^{-2}$) at $M_{\mathrm{halo}} \sim 10^{13} h^{-1}\,{\rm M}_{\odot}$ ($\sim 10^{11.5} h^{-1}\,{\rm M}_{\odot}$), and then becomes flat above these masses at $z=1$ ($z=4$).
The plateau continues up to the high mass end available in the simulation at both redshifts.
The mean occupation number of satellite AGNs (thin black lines of the right panels) increases with $M_{\mathrm{halo}}$ almost monotonically, dominates the total $\langle N \rangle$ at high mass haloes with $M_{\mathrm{halo}} \ga 10^{13.5} h^{-1}\,{\rm M}_{\odot}$ ($\ga 10^{12} h^{-1}\,{\rm M}_{\odot}$) at $z=1$ ($z=4$), and rises to $\sim 1$ ($\sim 0.1$).
In contrast to $\langle N_{\mathrm{cen}} \rangle$, in most cases the mean occupation number of satellite AGNs, $\langle N_{\mathrm{sat}} \rangle$, continues to rise with halo mass.
The different behaviours between the central and satellite HODs can be explained by the fact that a halo has either zero or one central AGN, while more than one satellite AGNs can occupy a single halo.
The form of the satellite HOD predicted by our model is similar to that of galaxies or dark matter subhaloes (e.g. \citealt{Kravtsov2004}; \citealt{Zheng2005}).
The trend in the satellite HOD is also similar to the theoretical results (\citealt{Chatterjee2012}) and the observational results (\citealt{Richardson2012}; \citealt{Shen2013}; \citealt{Chatterjee2013}), although the slope is steeper than the observations of \citet{Miyaji2011} and \citet{Allevato2012}.
In terms of the HODs of AGNs driven by disc instabilities, while $\langle N \rangle$ is small at $z=1$, it is above those by major mergers at $z=4$.
For the AGNs with $\log (L_{\rm X} / \mathrm{erg~s^{-1}}) = [42.5, 43.5]$, our result shows that minor mergers are the main triggering mechanism of AGNs except for high-z ($z \sim 4$) central AGNs.
We have predicted that the satellite AGNs contribute the HOD in high mass haloes.
\citet{Powell2018} also suggest that a significant fraction of AGNs is in satellites, using HOD models to the AGN clustering measurements.

Fig.~\ref{fig:hod_highlum_mod} shows the HODs of more luminous AGNs with $\log (L_{\rm X} / \mathrm{erg~s^{-1}}) = [43.5, 44.5]$.
The global trend is similar to the case of moderately luminous AGNs shown in Fig.~\ref{fig:hod_midlum_mod}.
The mean occupation number, $\langle N \rangle$, is, however, an order of magnitude smaller than in the case of moderately luminous AGNs at $z=1$; $\langle N_{\mathrm{cen}} \rangle$ and $\langle N_{\mathrm{sat}} \rangle$ rise to $\sim 10^{-2}$ and $\sim 10^{-1}$, respectively.
In this luminosity range, the contribution of major mergers is slightly higher than in the luminosity range $\log (L_{\rm X} / \mathrm{erg~s^{-1}}) = [42.5, 43.5]$, at $z=1$.
From Fig.~\ref{fig:hod_midlum_mod} and Fig.~\ref{fig:hod_highlum_mod}, we note that
the AGN host halo mass extends to low mass haloes ($M_{\mathrm{halo}} < 10^{11.0} h^{-1}\,{\rm M}_{\odot}$), in particular, for less luminous AGN and at higher redshift.
This indicates that it is important to resolve small mass haloes to obtain the AGN bias.
The high resolution cosmological $N$-body simulations we use in our semi-analytic model (\citealt{Ishiyama2015}), where the minimum halo mass is 8.79~$\times 10^9 h^{-1}\,{\rm M}_{\odot}$, enable us to analyse such low mass host haloes.
\citet{Georgakakis2019} have argued the same point by using an empirical model, with which they assign AGNs to the simulated DM haloes.

\subsubsection{Comparison with direct measurements of the HOD}
\label{subsubsec:comp_hod}

Next, we compare the prediction of our model with observations in Fig.~\ref{fig:hod_comparison}.
\citet{Leauthaud2015} selected AGNs in the redshift range $0.2 < z < 1$.
They focused on the moderately luminous AGNs and limited their sample to AGN with a rest-frame $0.5-10$~keV band luminosity in the range $10^{41.5} < L_{\rm X} < 10^{43.5}~\mathrm{erg~s^{-1}}$.
They also impose a lower limit on host galaxy stellar mass of $\log_{10} (M_{*}/{\rm M}_{\odot}) > 10.5$.
As a result, their sample has a mean redshift of $\langle z \rangle = 0.7$ and a mean-log X-ray luminosity of $\langle \log_{10}(L_{\rm X} / \mathrm{erg~s^{-1}}) \rangle = 42.7$.
To mimic their sample selection, we first estimate $L_{\rm X}(0.5-10~\mathrm{keV})$ from $L_{\rm X}(2-10~\mathrm{keV})$.
From previous studies (e.g. \citealt{Bianchi2009a, Bianchi2009b}), the mean value of the ratio, $L_{\rm X}(0.5-2~\mathrm{keV})/L_{\rm X}(2-10~\mathrm{keV})$, is $\sim1$ (see fig.~1 of \citealt{Bianchi2009b}).
Here, we simply regard $L_{\rm X}(0.5-10~\mathrm{keV})$ as twice $L_{\rm X}(2-10~\mathrm{keV})$ for all AGNs.
We use the model AGN sample at $z=0.69$ and adopt $\log (L_{\rm X}(2-10~\mathrm{keV}) / \mathrm{erg~s^{-1}}) = 42.0$ as the lower limit of $L_{\rm X}$, which roughly corresponds to the flux limit at $z\simeq0.7$ from fig.~1 of \citet{Leauthaud2015}.
Then, we limit our sample to AGNs with $\log (L_{\rm X}(0.5-10~\mathrm{keV}) / \mathrm{erg~s^{-1}}) = [42.0, 43.5]$ at $z=0.69$.
We also impose a lower limit of the stellar mass of the host galaxies, $\log_{10} (M_{*}/{\rm M}_{\odot}) > 10.5$, just as \citet{Leauthaud2015} did, for a fair comparison.
While \citet{Leauthaud2015} have analysed obscured (type-2) X-ray AGNs, we model all (type-1 and type-2) X-ray AGNs without obscuration for the X-ray AGNs.
For the comparison, we simply assume four obscured fraction, $f_{\mathrm{obs}} = 0.2$, 0.4, 0.6, and 1.0, and randomly select a sample from the all AGN sample.

Fig.~\ref{fig:hod_comparison} shows the resultant HODs for the four values of $f_{\mathrm{obs}}$ (see the figure legend for different line types).
We separately show the results for all, central, and satellite AGNs.
Broadly speaking, the shape of our predicted HODs is similar to the observation.
As expected, the model with $f_\mathrm{obs}=1$ overpredicts the observed HODs.
Then, by lowering this fraction to $\sim 0.2$, the model can fit the observed HODs. Remarkably, both the central and the satellite AGNs are fit quite well with this fraction.
However, this obscured fraction is smaller than the observational estimation (e.g. \citealt{Lusso2013}).
For example, \citet{Lusso2013} have shown that the obscured fraction ranges from 0.3 to 0.75.
As an exception, for the central AGNs in lower mass haloes ($M_{\mathrm{halo}}\sim 10^{12} {\rm M}_{\odot}$), the results with higher obscured fractions ($f_\mathrm{obs}= 0.6, 1$) agree better with the observed HOD.
This might reflect a trend of varying obscuration with luminosity and/or Eddington ratio.
Observational data shows that AGNs with lower luminosities and lower Eddington ratio have higher obscured fraction (\citealt{Lusso2013}; \citealt{Ricci2017}).
Although we could not find such trends in our mock AGNs, how the varying obscured fraction affects the AGN HOD would be interesting.
We postpone further detailed investigations for future studies.
We note that \citet{Leauthaud2015} do not correct the uncertainty from the sample incompleteness due to the flux limit.
The correction might increase the normalisation of their HOD such that the best-fitting obscured fraction is in a better agreement with the observational results.
Although there still remains detailed technical points for a quantitative and fully consistent comparison, as described above, the current level of agreement suggests that our model, which includes the AGN triggering through major and minor mergers and the gas accretion time-scale regulated by SMBH mass and accreted gas mass, explains the shape of the HOD of AGNs, especially the relative contribution from the satellites quite well (see the next subsection for more discussion on the satellite fraction).
We note again that the HOD is sensitive to the sample selection.
This should be more quantitatively studied in future.

\subsubsection{Satellite fraction}
\label{subsubsec:sat_frac}

\begin{figure}
	 \centering
	\includegraphics[width=0.9\columnwidth]{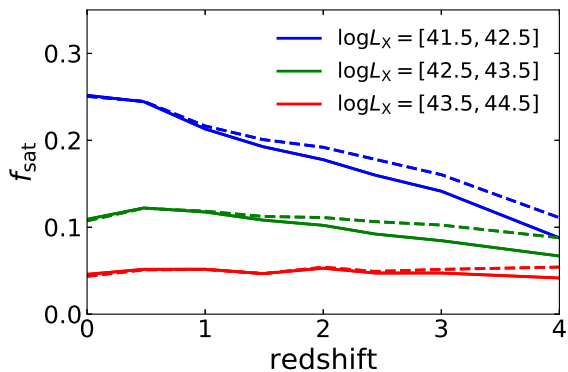}
    \caption{AGN satellite fraction for AGNs with luminosity ranges, $\log (L_{\rm X} / \mathrm{erg~s^{-1}}) =$ [41.5, 42.5] (blue), [42.5, 43.5] (green), and [43.5, 44.5] (red) as a function of redshift.
    Solid lines show the satellite fraction for AGNs in all DM haloes.
    Dashed lines show that for AGNs in haloes with mass $M_{\mathrm{halo}} \geq 10^{11} h^{-1}\,{\rm M}_{\odot}$.
    }
    \label{fig:sat_frac}
\end{figure}

The fraction of satellite AGNs to the total AGN population, the satellite fraction, can be inferred from the HOD modelling to the observations.
The observationally inferred satellite fraction is in the range $f_{\mathrm{sat}} \sim 0.001 - 0.1$ (\citealt{Starikova2011}; \citealt{Kayo2012}; \citealt{Richardson2012}; \citealt{Richardson2013}; \citealt{Shen2013}), while \citet{Leauthaud2015} infer a relatively high satellite fraction $f_{\mathrm{sat}} \sim 0.18$.
We show the predicted satellite fraction in our model as a function of redshift and AGN X-ray luminosity in Fig.~\ref{fig:sat_frac}.
Overall, the satellite fraction is higher for AGNs with lower luminosity.
For AGNs with $\log (L_{\rm X} / \mathrm{erg~s^{-1}}) = [43.5, 44.5]$, $f_{\mathrm{sat}} \sim 0.05$ and does not depend on redshift significantly.
For AGN with $\log (L_{\rm X} / \mathrm{erg~s^{-1}}) = [42.5, 43.5]$, $f_{\mathrm{sat}}$ weakly depends on redshift and $f_{\mathrm{sat}} = 0.12$ at $z = 0.48$.
For AGN with $\log (L_{\rm X} / \mathrm{erg~s^{-1}}) = [41.5, 42.5]$, $f_{\mathrm{sat}}$ significantly depends on redshift and reaches $\sim 0.25$ at $z=0$.
The satellite fraction in our fiducial model with the luminosity range matched to \citet{Leauthaud2015} is 0.23, roughly consistent with their conclusion.
In addition, we investigate the effect of the minimum halo mass on the satellite fraction.
The dashed lines in Fig.~\ref{fig:sat_frac} show the fraction for AGNs in haloes with mass $M_{\mathrm{halo}} \geq 10^{11} h^{-1}\,{\rm M}_{\odot}$.
We find that the the minimum halo mass cut does not affect the result at $z \la 1$, however, at $z>1$, $f_{\mathrm{sat}}$ increases for AGNs with $\log (L_{\rm X} / \mathrm{erg~s^{-1}}) < 43.5$.

\subsection{The origin of satellite AGNs}

In our model, satellite AGNs make a non-negligible contribution to the total number of AGNs and affect the host halo mass distribution of AGNs.
In this subsection, we examine the origin of the satellite AGNs.
As shown in the right panels of Figs~\ref{fig:hod_midlum_mod} and \ref{fig:hod_highlum_mod}, in our model, minor mergers are the main triggering mechanism for satellite AGNs, and satellite AGNs driven by major mergers and disc instabilities constitute a subdominant population.
These satellite AGNs have two distinct origins.
The first is an AGN which is activated when the host galaxy was a central galaxy, and is accreted into a larger halo after that while the activity is ongoing.
The second is an AGN which is triggered by a merger between satellite galaxies, which is implemented as a random collision between satellite galaxies in the same halo in our model (\citealt{Makiya2016}; \citealt{Shirakata2019}).
We call the latter `sat-sat merger AGN'.

We find that the fraction of the sat-sat merger AGNs to the total satellite AGNs, $f_{\mathrm{sat-sat}}$, increases with cosmic time for AGNs with the three luminosity ranges.
At $z=3$, $f_{\mathrm{sat-sat}} \sim 0.1$ for AGNs with $\log (L_{\rm X} / \mathrm{erg~s^{-1}}) = [43.5, 44.5]$, and $\sim 0.3$ for those with $\log (L_{\rm X} / \mathrm{erg~s^{-1}}) = [41.5, 42.5]$ and $[42.5, 43.5]$.
At $z\la2$, the sat-sat merger AGNs become the predominant population in the satellite AGNs with $\log (L_{\rm X} / \mathrm{erg~s^{-1}}) = [41.5, 42.5]$ and $[42.5, 43.5]$.
At $z=0$, $f_{\mathrm{sat-sat}}$ reaches 0.6--0.8 for all the luminosity ranges.
Thus, our model predicts that the mergers between satellite galaxies are a key process for the triggering mechanism of satellite AGNs contributing significantly to the satellite HOD and the satellite fraction at low redshift.

\subsection{Comparison between two bias measurements}
\label{subsec:b_comparison}

\begin{figure}
	 \centering
	\includegraphics[width=0.9\columnwidth]{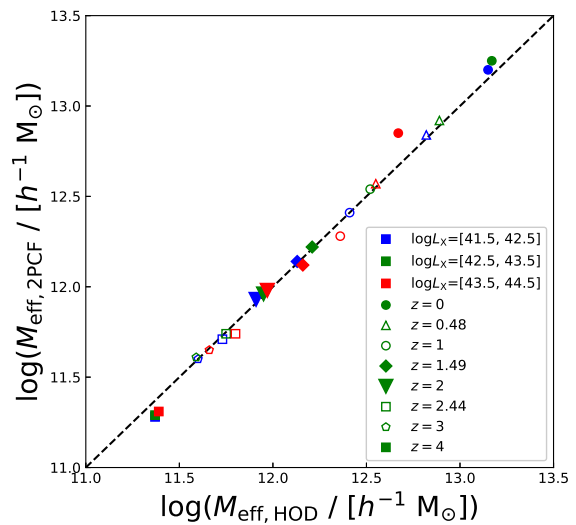}
    \caption{Relation between the effective halo mass obtained with the 2PCF, $M_{\mathrm{eff,2PCF}}$, and that obtained with Eqn.~\ref{eq:bias_hod}, $M_{\mathrm{eff,HOD}}$, for AGNs with luminosity ranges, $\log (L_{\rm X} / \mathrm{erg~s^{-1}}) =$ [41.5, 42.5] (blue), [42.5, 43.5] (green), and [43.5, 44.5] (red) at various redshifts.
    }
    \label{fig:meff_comparison}
\end{figure}

The effective bias of objects can be calculated using the following equation (\citealt{Baugh1999}; \citealt{Enoki2003}; \citealt{Fanidakis2013a}):
\begin{equation}
\label{eq:bias_hod}
b_{\mathrm{eff}} = \frac{\int b(M_{\mathrm{halo}}) \langle N_{\mathrm{AGN}} (M_{\mathrm{halo}}) \rangle n(M_{\mathrm{halo}}) d\log M_{\mathrm{halo}} } {\int \langle N_{\mathrm{AGN}} (M_{\mathrm{halo}}) \rangle n(M_{\mathrm{halo}}) d\log M_{\mathrm{halo}}},
\end{equation}
where $b(M_{\mathrm{halo}})$, $\langle N_{\mathrm{AGN}} (M_{\mathrm{halo}}) \rangle$, and $n(M_{\mathrm{halo}})$ are the DM halo bias, the mean number of AGNs, and the number density of DM haloes as a function of halo mass, respectively.
The equation assumes that the bias of the haloes hosting the objects only depends on the mass of the haloes and is independent on scale.
This method has an advantage of deriving the effective bias of objects without spatial statistics, the 2PCF.
Many previous studies use this equation to derive the bias of objects (e.g. \citealt{Enoki2003}; \citealt{Fanidakis2013a}).
\citet{Enoki2003} have derived, for the first time, the quasar bias by using this equation.
It is, however, uncertain whether the effective bias is consistent with the bias derived from the square root of the ratio of the 2PCF of the AGNs to that of DM, which we use in this paper (Eqn.~\ref{eq:bias}).

Here, we investigate the difference between the AGN bias derived from the spatial distribution of our model AGNs based on the cosmological $N$-body simulations and that by using Eqn.~\ref{eq:bias_hod}.
For $b(M_{\mathrm{halo}})$, we use the formula from \citet{Tinker2010}.
In order to clarify how the difference of the AGN bias causes the difference of $M_{\mathrm{halo,eff}}$, we present the relation between two $M_{\mathrm{halo,eff}}$ corresponding to two AGN bias; that obtained by using the 2PCF, $M_{\mathrm{eff,2PCF}}$, and that obtained by using Eqn.~\ref{eq:bias_hod}, $M_{\mathrm{eff,HOD}}$, in Fig.~\ref{fig:meff_comparison}.
Here, we use the 2PCF at 8~$h^{-1}$~Mpc for the calculation of $M_{\mathrm{eff,2PCF}}$.
The two effective halo masses are similar and there is small difference.
This result indicates that the bias of AGNs, at least at large-scale, can be estimated by using Eqn.~\ref{eq:bias_hod} fairly well.
Furthermore, when using Eqn.~\ref{eq:bias_hod}, we assume that the halo bias depends only on the halo mass.
Thus, this result may also imply that AGN host haloes are not affected by the halo assembly bias, that is, the bias related to the halo assembly history rather than the halo mass (e.g. \citealt{Gao2005MNRAS.363L..66G}; \citealt{Dalal2008}).

\section{Summary and Discussion}
\label{sec:discussion}

We have predicted the AGN host halo mass and clustering by using our latest semi-analytic model, $\nu^2$GC.
This model includes a new prescription for the gas accretion time-scale that is required to match the faint end of the observed luminosity function of X-ray AGNs (\citealt{Shirakata2019}).
We have found that the gas accretion time-scale on to SMBHs also plays a significant role in the clustering properties of AGNs.
In the model with long gas accretion time-scales (our fiducial model), there is a significant fraction of AGNs with $40 \la \log (L_{\rm X} / \mathrm{erg~s^{-1}}) \la 43$ which reside in high mass haloes with $M_{\mathrm{halo}} \ga 10^{13} h^{-1}\,{\rm M}_{\odot}$ (Fig.~\ref{fig:lx_mhalo_4panel}).
This feature is not seen in the model without the long accretion time-scales (galmodel).
We have derived the effective halo masses of AGNs in the two models from their clustering bias factors of and compared them.
The effective halo mass from our fiducial model is larger than that from the galmodel for the luminosity range, $41.5 \leq \log (L_{\rm X} / \mathrm{erg~s^{-1}}) \leq 43.5$, at low redshift (Fig.~\ref{fig:z_meff_4panel}).
In addition to the effective halo mass of AGNs we have also investigated the 2PCF using our cosmological $N$-body simulation.
Our simulation has an unprecedentedly large box size and high mass resolution, enabling the prediction of the auto-correlation function of rare objects, such as AGNs, with high accuracy.
We have shown a clear difference in the 2PCF between the fiducial model and the galmodel (Fig.~\ref{fig:2pcf_obs}).
Our results suggest that the 2PCF provides an independent constraint on the gas accretion time-scale complementary to the AGN luminosity function.

In our model, we assume that SMBHs grow through the cold gas accretion associated with starbursts during galaxy major/minor mergers and disc instabilities.
Our results suggest that the observed clustering properties and the inferred host halo mass of X-ray AGNs can be explained by the model with these processes.
As seen in Fig.~\ref{fig:z_meff_4panel}, the effective halo mass of moderately luminous X-ray AGNs reaches  $\sim 10^{13} h^{-1}\,{\rm M}_{\odot}$ at $z \la 1$.
Also, we show that the effective halo mass,
$M_{\mathrm{halo, eff}}$, has a negative dependence on the AGN luminosity (Fig.~\ref{fig:lumx_meff_3panel}).
This trend is consistent with observations; less luminous X-ray selected AGNs tend to be in more massive haloes compared with more luminous optically selected quasars.
While the median of the host halo mass distribution is less than $10^{13} h^{-1}\,{\rm M}_{\odot}$, the distribution has a positive skewness with a heavy tail towards the massive end (Fig.~\ref{fig:m_halo_dist}).
This feature is considered to contribute the larger effective halo mass (Fig.~\ref{fig:mmed_meff}) compared with the median value in the distribution.
This skewness and the tail are consistent with the results of a semi-empirical model (\citealt{Georgakakis2019}) and the observation (\citealt{Leauthaud2015}).
We have also compared the predicted 2PCF and HOD of X-ray AGNs with observations (Figs~\ref{fig:2pcf_obs} and \ref{fig:hod_comparison}).
Our results are consistent with the current measurements of the 2PCF from observations and halo occupation statistics based on the observation with the obscured fraction treated as a free parameter (\citealt{Leauthaud2015}).
This supports the HOD modelling based on the observations, in which satellite HODs monotonically increases with halo mass (\citealt{Richardson2012}; \citealt{Shen2013}).
Furthermore, we have shown that the predicted AGN satellite fraction (Fig.~\ref{fig:sat_frac}) is also consistent with the current observations.

One interesting trend in Figs~\ref{fig:hod_midlum_mod} and \ref{fig:hod_highlum_mod} is that the relative importance of different channels for the AGN trigger can vary quite significantly over cosmic time.
While the dominant mechanism is found to be minor mergers for all the redshifts investigated here, the disc instabilities are not negligible at high redshifts.
Therefore, a more accurate measurement of the 2PCF at those redshifts would be important to further test the validity of our model.
Also, there are some other mechanisms which are not considered in our model, such as hot-halo mode AGN activity.
\citet{Fanidakis2013a} have concluded that X-ray AGNs are triggered by the hot-halo mode gas accretion, which can explain the massive host halo for the X-ray AGNs.
Although they consider the different mechanism, interestingly, they also predict the negative dependence of the effective halo mass on the X-ray luminosity of AGNs near the high luminosity end.
However, the maximum effective halo mass of their model is $\sim$0.4~dex larger than our prediction at $z\sim0$.
\citet{Allevato2011} have estimated the effective halo mass significantly larger than those of \citet{Starikova2011} and \citet{Mountrichas2016} at $z\sim1$.
If the former is the case, it might suggest that the hot-halo mode AGN activity plays an important role in the host halo mass.
To further test which is more plausible, a precise observational determination of the 2PCF at low redshifts would be important.

The main triggering mechanism of the SMBH growth remains unclear, as mentioned above.
Here, we consider the possibilities that further measurements including the 2PCF clarify the main driver of the SMBH growth.
In our model, minor mergers with long gas accretion time-scales are the main trigger for low luminosity AGNs ($\lesssim 10^{43} \mathrm{erg~s^{-1}}$).
In contrast, in \citet{Fanidakis2012}, hot-halo mode AGNs dominate the low luminosity end.
While both the two scenarios are consistent with the current clustering measurements, detailed shapes of the 2PCF may be different.
This is because the two scenarios can predict distinct AGN host galaxies.
\citet{Fanidakis2013a, Fanidakis2013b} have claimed that the low luminosity AGNs are hosted by elliptical galaxies in galaxy groups and clusters.
These host galaxies are expected to be central galaxies in the haloes.
On the other hand, our model AGNs in the same luminosity range reside in both central and satellite galaxies.
As a result, the two scenarios predict different shapes of the 2PCF, in particular, the small-scale clustering.
Therefore, a more accurate measurement of the 2PCF over a wide separation range is useful to distinguish the two scenarios.
The accurate clustering measurement will be provided by future surveys such as eROSITA (\citealt{Merloni2012}).
Moreover, the properties of the host galaxies, such as the morphology, colour, and star formation rate, are also key quantities to distinguish these models.
In addition to the hot-halo mode, in \citet{Fanidakis2012}, disc instabilities are the major triggering mechanism of the SMBH growth and AGN activity for higher luminosities, in contrast to our model with minor mergers as the main driver of the SMBH growth.
Further studies of the morphology of AGN host galaxies could distinguish the two triggering mechanisms: minor mergers and disc instabilities (e.g. \citealt{Rosario2015}; \citealt{Mechtley2016}; \citealt{Marian2019}).

We have shown that the gas accretion time-scale plays a significant role in the clustering properties of AGNs.
Specifically, the long accretion time-scale for low luminosity AGNs and at low redshifts has significant contribution to the massive host halo mass.
This means that as well as the process triggering the gas supply to the galactic centre, the gas accretion physics regulating the accretion time-scale at the centre is important for understanding the AGN statistics and the co-evolution of galaxies and SMBHs.
Our finding is consistent with that of \citet{Bonoli2009}, which have also shown that AGN light curve models affect the correlation length of the AGN clustering, while they have not addressed the clustering of X-ray AGNs.
Using similar models, \citet{Marulli2009} have shown that the luminosity dependence of the correlation length of X-ray AGNs is sensitive to the AGN light curve models, although they have underpredicted the correlation length compared with observations and have not investigated the host halo masses of X-ray AGNs.
Based on the outcomes above, future observational measurements of the AGN clustering with larger samples, which enables us to investigate the luminosity dependence of the AGN clustering more accurately, can provide a constraint on the gas accretion time-scale on to SMBHs and the AGN light curves.

The AGN flickering (\citealt{Schawinski2015}) may affect the statistics of AGN populations.
The AGN variability manifests itself in observations as the light echo (\citealt{Lintott2009}; \citealt{Keel2012AJ....144...66K, Keel2012MNRAS.420..878K}).
Hydrodynamic simulations of SMBH accretion have shown that the short time-scale variability is driven by the stochastic accretion of gas clumps on to the central BH (\citealt{Novak2011}; \citealt{Gabor2013}).
This variability can cause the additional scatter in the relation between the AGN luminosity and the host halo mass and may affect the AGN bias and effective halo mass.

We have investigated the effects of the AGN flickering on the 2PCF of AGNs.
We change the AGN luminosity assuming the log-normal distribution with $\sigma_{\log L}$ between 0.3 and 0.5~dex per luminosity to mimic the impact of flickering following the work by \citet{Rosas-Guevara2016}.
In the model with 0.5~dex scatter, the amplitude of the 2PCF is slightly larger than that of the fiducial model, though this is subdominant compared to the statistical uncertainties of the 2PCF and thus does not affect our main conclusion significantly.
This is because the scatter in $M_{\mathrm{halo}}-L_{\mathrm{X}}$ relation is already sufficiently large in our fiducial model, so the effect of the additional scatter is negligible.
We conclude that the AGN flickering does not have a large effect on the 2PCF in our model.
However, to investigate the AGN flickering effect by using a more physically motivated model would be meaningful.


As we have analysed in this paper, the HODs of AGNs can provide a further constraint on the AGN formation models.
While some previous studies have inferred the AGN HOD by using the 2PCFs of AGNs, there are few attempts to obtain the HOD directly.
In addition to the HOD, other observations are also expected to be a key constraint on the theoretical models.
Recent observations have investigated the dependence of AGN clustering on other AGN and host galaxy properties such as SMBH mass, the Eddington ratio (\citealt{Krumpe2015}), stellar mass, and star-formation activity (\citealt{Mountrichas2019}; see also \citealt{Viitanen2019arXiv190607911V}).
These measurements provide alternative constraints on AGN formation models.
Further studies to constrain the theoretical models using these additional measurements will be done in the future.

\section*{Acknowledgements}

We are grateful to the anonymous referee for providing constructive comments.
We thank M. Oguri, C. Hikage, H. Ikeda, T. Miyaji, T. Kirihara, K. Wada and N. Yoshida for useful comments and discussion.
This research was supported by World Premier International Research Center Initiative (WPI), MEXT, Japan and by
MEXT as ``Priority Issue on Post-K computer'' (Elucidation of the Fundamental Laws and Evolution of the Universe), MEXT as ``Program for Promoting Researches on the Supercomputer
Fugaku'' (Toward a unified view of the universe: from large scale
structures to planets), and JICFuS.
This study has been funded by MEXT/JSPS KAKENHI Grant Number 17H02867, 18H05437, 20H01950 (MN), JP17K14273, JP19H00677 (TN), JP17H04828, JP17H01101, JP18H04337 (TI), 18H04333, 19H01931 (T. Okamoto), by Yamada Science Foundation, MEXT HPCI STRATEGIC PROGRAM, and by Japan Science and Technology Agency CREST JPMHCR1414.

\section*{Data Availability}

The data underlying this article will be shared on reasonable request to the corresponding author.



\bibliographystyle{mnras}
\bibliography{ref} 








\bsp	
\label{lastpage}
\end{document}